\documentclass[twocolumn,prl,aps,superscriptaddress,amsfonts,showpacs,longbibliography,notitlepage,nofootinbib]{revtex4-2}
\usepackage[colorlinks=true,urlcolor=blue,citecolor=blue,linkcolor=blue]{hyperref}
\usepackage[dvipsnames]{xcolor}
\usepackage{bm}
\usepackage{bbm}
\usepackage{braket}
\usepackage[utf8]{inputenc}
\usepackage[T1]{fontenc}
\usepackage{amsmath}
\usepackage{mathrsfs}
\usepackage{graphicx}
\usepackage{booktabs}       
\usepackage{multirow}       
\usepackage{algorithm}      
\usepackage{algpseudocode}  
\usepackage{longtable}

\renewcommand{\vec}[1]{\boldsymbol{#1}}
\newcommand{\Eq}[1]{Eq.~(\ref{#1})}
\newcommand{\Fig}[1]{Fig.~\ref{#1}}

\newcommand{\Table}[1]{Table.~\ref{#1}}
\newcommand{\Tr}{\mathrm{Tr}}

\algnewcommand\algorithmicforeach{\textbf{for each}}
\algdef{S}[FOR]{ForEach}[1]{\algorithmicforeach\ #1\ \algorithmicdo}

\renewcommand{\section}[1]{\textit{#1---}}

\begin{document}
\title{Deep Variational Free Energy Calculation of Hydrogen Hugoniot}

\author{Zihang Li}
\affiliation{Beijing National Laboratory for Condensed Matter Physics and Institute of Physics, \\Chinese Academy of Sciences, Beijing 100190, China}
\affiliation{University of Chinese Academy of Sciences, Beijing 100049, China}

\author{Hao Xie}
\email{qwexiehao@gmail.com}
\affiliation{Beijing National Laboratory for Condensed Matter Physics and Institute of Physics, \\Chinese Academy of Sciences, Beijing 100190, China}
\affiliation{Department of Astrophysics, University of Zürich, Winterthurerstrasse 190, 8057 Zürich, Switzerland}

\author{Xinyang Dong}
\email{dongxy@iphy.ac.cn}
\affiliation{Beijing National Laboratory for Condensed Matter Physics and Institute of Physics, \\Chinese Academy of Sciences, Beijing 100190, China}

\author{Lei Wang}
\email{wanglei@iphy.ac.cn}
\affiliation{Beijing National Laboratory for Condensed Matter Physics and Institute of Physics, \\Chinese Academy of Sciences, Beijing 100190, China}

\date{\today}

\begin{abstract}
We develop a deep variational free energy framework to compute the equation of state of hydrogen in the warm dense matter region. This method parameterizes the variational density matrix of hydrogen nuclei and electrons at finite temperature using three deep generative models:
a normalizing flow model for the Boltzmann distribution of the classical nuclei, an autoregressive transformer for the distribution of electrons in excited states, and a permutational equivariant flow model for the unitary backflow transformation of electron coordinates in Hartree-Fock states. By jointly optimizing the three neural networks to minimize the variational free energy, we obtain the equation of state and related thermodynamic properties of dense hydrogen for the temperature range where electrons occupy excited states. 
We compare our results with other theoretical and experimental results on the deuterium Hugoniot curve, aiming to resolve existing discrepancies. 
Our results bridge the gap between the results obtained by path-integral Monte Carlo calculations at high temperature and ground-state electronic methods at low temperature, thus providing a valuable benchmark for hydrogen in the warm dense matter region.
\end{abstract}

\maketitle

\section{Introduction}
Hydrogen is the most abundant element in the universe. A full understanding of the hydrogen phase diagram under extreme conditions~\cite{RevModPhys.84.1607, Bonitz_2024} is important for many applications ranging from planetary science~\cite{KERLEY197278, Militzer_2008, Baraffe_2010, miguel_jupiter_2016} to inertial confinement fusion~\cite{10.1063/1.871025, 10.1063/1.4934714, 10.1063/1.4917477}. In particular, in the warm dense matter (WDM) region~\cite{Falk_2018, 10.1063/1.5143225, graziani_frontiers_2014, vorberger2025roadmapwarmdensematter, PhysRevLett.129.066402}, hydrogen undergoes the molecular-to-atomic transition~\cite{PhysRevE.63.066404, moldabekov2023bound}, approaches metallization~\cite{knudson_direct_2015}, and exhibits a delicate interplay of quantum and thermal effects in dense plasmas~\cite{PhysRevE.61.3470, PhysRevLett.76.1240}.
A thorough understanding of this regime is crucial for accurately modeling hydrogen-rich giant planets (such as Jupiter), and for optimizing inertial-confinement-fusion implosions, whose efficiency hinges on the chosen compression pathway.

The Hugoniot curve~\cite{1967pswh.book}, obtained by tracing the locus of shock-compressed states during dynamic loading of hydrogen, offers a crucial experimental benchmark for constraining hydrogen's equation of state (EOS) in the WDM region. 
Experimental studies commonly focus on the deuterium Hugoniot curve because deuterium's higher mass density facilitates both the attainment and accurate diagnosis of extreme shock-compressed states~\cite{RevModPhys.84.1607}. 
Landmark experiments using gas guns~\cite{10.1063/1.439955, 10.1063/1.445938, PhysRevA.27.608, PhysRevB.52.15835}, lasers~\cite{PhysRevLett.78.483, collins_measurements_1998, PhysRevLett.87.165504, PhysRevLett.90.035505, 10.1063/1.1778164, PhysRevB.79.014112, PhysRevLett.103.225501, PhysRevB.83.054117, PhysRevLett.122.255702, PhysRevB.103.134107}, explosions~\cite{belov_shock_2002, boriskov_shock-wave_2003, grishechkin_experimental_2004, PhysRevB.71.092104}, and pulsed power facilities~\cite{10.1063/1.1891746, PhysRevLett.87.225501, PhysRevB.69.144209, PhysRevB.78.144107, PhysRevB.91.224105, PhysRevLett.118.035501} have measured the deuterium Hugoniot up to megabar pressures, revealing anomalies such as unexpectedly high compressibility near 50 GPa. These results have not been fully agreed upon~\cite{PhysRevLett.89.165502, Bonitz_2024}, and have sparked debates over dissociation mechanisms and electronic structure changes, yet experimental data remain sparse at higher temperatures due to diagnostic challenges. 

Complementary theoretical approaches have been employed to tackle the deuterium Hugoniot curve challenge, 
including path-integral Monte Carlo (PIMC)~\cite{militzer1999characterization, PhysRevLett.85.1890, PhysRevLett.87.275502, PhysRevE.70.057401, filinov_calculation_2005, PhysRevLett.104.235003, PhysRevB.84.224109, khairallah2011pathintegralcalculationshydrogen}, coupled electron-ion Monte Carlo (CEIMC)~\cite{PhysRevLett.115.045301, PhysRevB.102.144108}, density functional theory (DFT)~\cite{PhysRevB.61.1, PhysRevB.68.064204, PhysRevB.69.014101, PhysRevB.77.184201, 10.1063/1.3467969, PhysRevB.83.094101, MORALES20125, 10.1063/1.4821839, PhysRevB.99.214110}, machine learning based methods~\cite{PhysRevB.110.L041107, 10.1063/5.0254638}, and various alternative techniques~\cite{PhysRevE.61.3470, PhysRevB.56.5164, PhysRevB.58.669, 1995ApJS...99..713S, PhysRevE.104.045304, Knaup_2003, 10.1063/1.480939}.
Databases such as SESAME \cite{KERLEY197278} and FPEOS \cite{PhysRevE.103.013203} compile EOS data generated by different methods, providing standardized references for both experimental and computational studies.
Nevertheless, these collective efforts remain insufficient. For example, restricted PIMC (RPIMC) may suffer from uncontrolled fixed-node error at low temperatures and high density~\cite{PhysRevLett.85.1890}, while CEIMC treats electrons in the ground state thus may become unreliable at finite temperatures~\cite{PhysRevLett.115.045301, PhysRevB.102.144108}. To date, there is still a lack of reliable quantum many-body calculations that can provide consistent predictions across the whole temperature range of the Hugoniot curve. Within the framework of quantum many-body methods, the handshake between finite-temperature and ground state electronic calculations along the Hugoniot curve remains to be established.

In this Letter, we combine the deep variational free energy method~\cite{PhysRevLett.131.126501, rbsg-r7hx}, previously 
applied to ground-state dense hydrogen, with the neural canonical transformation approach~\cite{JML-1-1,10.21468/SciPostPhys.14.6.154} to compute the excited-state deuterium equation of state and its Hugoniot curve in the temperature regime where electrons occupy excited states, in order to achieve the handshake for the deuterium Hugoniot curve from the finite-temperature side. This regime is particularly challenging for existing methods: ground-state approaches become unreliable when thermal electron excitations are significant, while PIMC methods encounter the fermion sign problem at lower temperatures. We aim to elucidate the phase behavior and provide a reliable EOS for deuterium to serve as a benchmark for future shock experiments and astrophysical modeling. The source code, raw data, and network checkpoints are openly available in the GitHub repositories~\cite{sourcecode,hqc} and HuggingFace repository~\cite{huggingface}.

\section{Method}
Consider a system of $N$ deuterium atoms in a periodic cubic cell of side length $L = (4\pi N/3)^{1/3} r_s a_0$, where $r_s$ is the dimensionless Wigner-Seitz parameter and $a_0$ is the Bohr radius, the Hamiltonian reads
\begin{align}
    \mathcal{H} = &-\frac{1}{2}\sum_{i=1}^N\nabla_i^2+\sum_{I<J}^N\frac{1}{\left|\vec{s}_I-\vec{s}_J\right|} \nonumber \\
    &-\sum_{i,I=1}^N\frac{1}{\left|\vec{r}_i-\vec{s}_I\right|}+\sum_{i<j}^N\frac{1}{\left|\vec{r}_i-\vec{r}_j\right|} \, ,
    \label{eq:Hamiltonian}
\end{align}
where $\vec{s} \equiv \{\vec{s}_1,\cdots,\vec{s}_N\}$ and $\vec{r} \equiv \{\vec{r}_1,\cdots,\vec{r}_N\}$ denote the coordinates of the nuclei and electrons, respectively.
Within the temperature range relevant to deuterium Hugoniot experiments (approximately $10^3\sim10^5$K), the nuclei can be treated as classical particles. 
Therefore, whether the nuclei are protons or deuterons only affects the entropy of classical nuclei~\cite{PhysRevLett.131.126501}. 
We focus on deuterium in the following discussion to benchmark with experimental and other computational results. 
The electrons need to be treated as quantum degenerate at finite temperature since the temperature remains below their Fermi temperature ($\sim 10^5$K), but a non-negligible fraction would occupy excited states when the temperature exceeds $10^4$K~\cite{PhysRevB.102.144108}. 
To properly account for the finite temperature effects on both nuclei and electrons, we search for variational density matrix of the following form
\begin{align}
    \rho = \int d{\vec{s}}\ p(\vec{s})\sum_{\vec{k}} p(\vec{k}|\vec{s})\ket{\vec{s},\Psi_{\vec{s},\vec{k}}}\bra{\vec{s},\Psi_{\vec{s},\vec{k}}} \, .
    \label{eq:DensityMatrix}
\end{align}
In \Eq{eq:DensityMatrix}, $p(\vec{s})$ represents the nuclear Boltzmann distribution, and the index $\vec{k}$ is used to label the (principally complete) set of electronic basis states under given nuclear coordinates $\vec{s}$.
$p(\vec{k}|\vec{s})$ represents the electronic excitation distribution in these states and $\Psi_{\vec{s},\vec{k}}(\vec{r})$ are the corresponding wave functions.
The variational free energy $F\left[\rho\right] = \frac{1}{\beta}\Tr\left(\rho\ln\rho\right) + \Tr\left(\rho \mathcal{H}\right)$ of the system then reads
\begin{align}
    F 
    = \mathop{\mathbb{E}}_{\vec{s}\sim p(\vec{s})} \bigg\{\frac{1}{\beta}\ln{p(\vec{s})} 
    &+ \mathop{\mathbb{E}}_{\vec{k}\sim p(\vec{k}|\vec{s})} \bigg[\frac{1}{\beta}\ln{p(\vec{k}|\vec{s})} \label{eq:FreeEnergy} \\
    &+ \mathop{\mathbb{E}}_{\vec{r}\sim \left|\Psi_{\vec{s},\vec{k}}(\vec{r})\right|^2} \bigg(
    \frac{\mathcal{H}\Psi_{\vec{s},\vec{k}}(\vec{r})}{\Psi_{\vec{s},\vec{k}}(\vec{r})} 
    \bigg)\bigg]\bigg\} \, , \nonumber 
\end{align}
where the first two terms are related to the nuclear and electronic entropy, respectively. The last term is the internal energy of the system.

\begin{figure}
    \centering
    \includegraphics[width=0.9\linewidth]{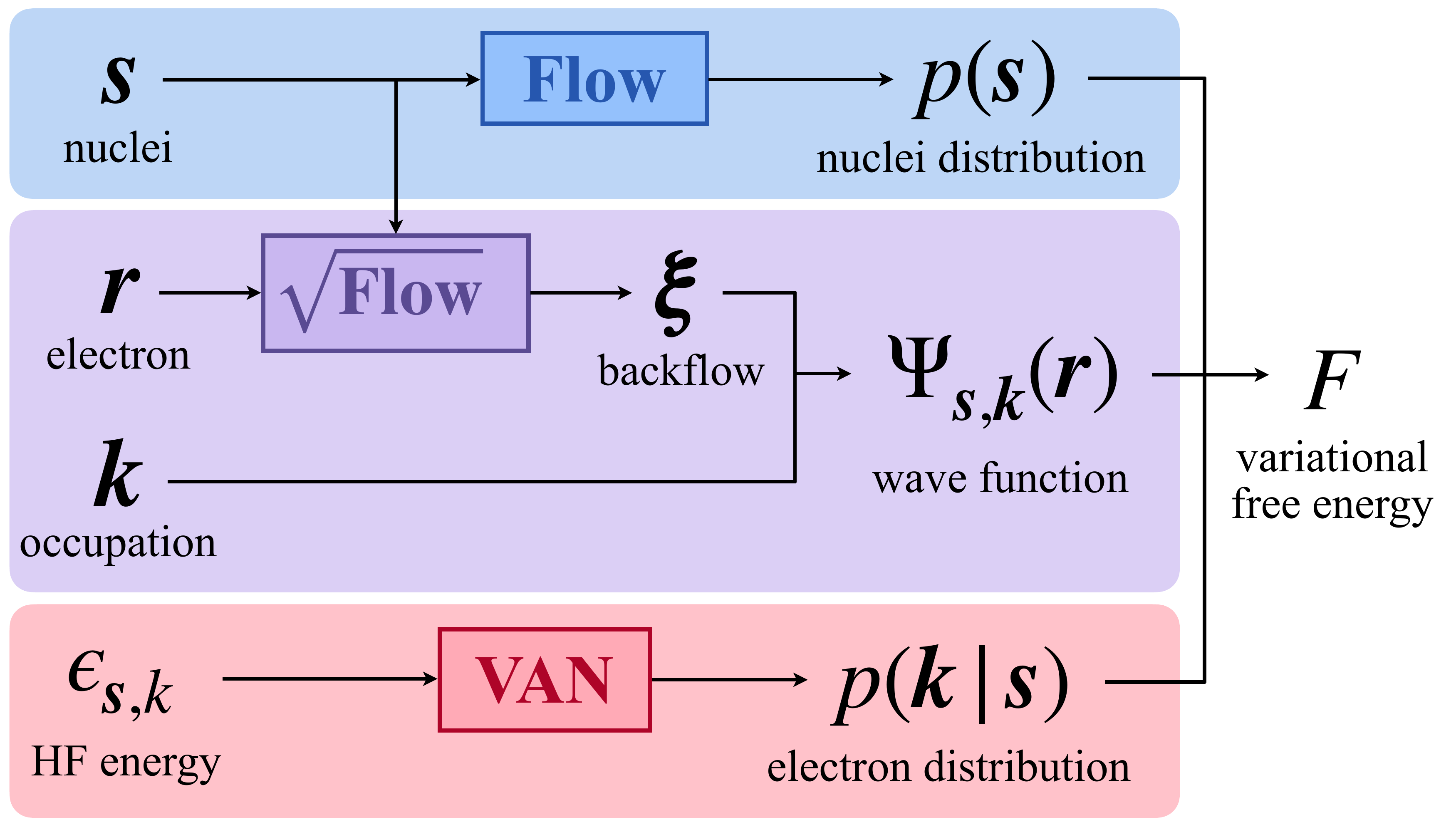}
    \caption{Computational graph for the finite electron temperature variational free energy calculation. The model consists of three trainable components: a normalizing flow \Eq{eq:boltzmann} for the nucleus Boltzmann distribution $p(\vec{s})$ in the blue part; a variational autoregressive network \Eq{eq:transformer1}, \Eq{eq:transformer2a}, and \Eq{eq:transformer2b} for the electron excitation distribution $p(\vec{k}|\vec{s})$ in the red part; and a quantum flow for the electron wave functions $\Psi_{\vec{s},\vec{k}}(\vec{r})$ shown in \Eq{eq:wavefunction} in the purple part. 
    We jointly optimize the three neural networks to minimize the variational free energy $F$ in \Eq{eq:FreeEnergy}. 
    }
    \label{fig:method}
\end{figure}

By parameterizing the nuclear Boltzmann distribution $p(\vec{s})$, electronic excitation distribution $p(\vec{k}|\vec{s})$, and electron wave function $\Psi_{\vec{s},\vec{k}}(\vec{r})$ by three generative models, one can then minimize \Eq{eq:FreeEnergy} to obtain thermodynamic properties of the system. All three generative models employed in this calculation have tractable normalization factors, which is crucial for estimating the entropy term in the variational free energy. A major difference between \Eq{eq:FreeEnergy} and the ground-state variational free energy expression employed in Refs.~\cite{PhysRevLett.131.126501, rbsg-r7hx} is that the present approach incorporates both electronic ground and excited states to account for thermal effects on electrons in the WDM region. 
The computational graph of the variational free energy method is shown in \Fig{fig:method}.

The Boltzmann distribution $p(\vec{s})$ of the nuclei is represented by a flow model, which involves a bijective map between the real coordinates $\vec{s}$ and a set of uniformly distributed "collective" coordinates $\vec{\zeta}$. The corresponding probability density reads
\begin{align}
    p(\vec{s}) = \frac{1}{L^{3N}}\left|\det\left(\frac{\partial \vec{\zeta}}{\partial\vec{s}}\right)\right| \, . \label{eq:boltzmann}
\end{align}
To model a complete set of electronic basis states $\ket{\Psi_{\vec{s},\vec{k}}}$ under given nuclear coordinates $\vec{s}$, we adopt the idea of neural canonical transformation proposed in Refs.~\cite{JML-1-1, 10.21468/SciPostPhys.14.6.154}. Specifically, we unitarily transform the set of Hartree-Fock (HF) Slater determinant states by applying a ($\vec{s}$-dependent) bijective map $\vec{r} \leftrightarrow \vec{\xi}$ to the electron coordinates. The HF states are computed using a batched HF solver on GTH-DZV basis set, which is sufficient for temperatures up to $62\,500$ K in this study. The resulting many-body wave function reads
\begin{align}
    &\Psi_{\vec{s},\vec{k}}(\vec{r}) \nonumber \\
    &= \det\left[\psi_{\vec{s},k_i^{\uparrow}}(\boldsymbol{\xi}_j^{\uparrow})\right]\det\left[\psi_{\vec{s},k_i^{\downarrow}}(\boldsymbol{\xi}_j^{\downarrow})\right]\left|\det\left(\frac{\partial\boldsymbol{\xi}}{\partial \vec{r}}\right)\right|^{\frac{1}{2}}.
    \label{eq:wavefunction}
\end{align}
In this context, the state index $\vec{k} \equiv \{k_1^\uparrow, \cdots, k_{N/2}^\uparrow,$ $k_1^\downarrow, \cdots, k_{N/2}^\downarrow\}$ is naturally labeled by $N$ occupied HF single-particle orbitals $\psi_{\vec{s},k_i^\sigma}(\boldsymbol{\xi}_j^\sigma)$, where $\sigma \in \{\uparrow, \downarrow\}$ denotes the electron spin. \Eq{eq:wavefunction} is similar in spirit to the backflow transformation, and the transformed coordinates $\vec{\xi}$ can be physically understood as quasiparticles. The difference lies in the additional Jacobian determinant factor in \Eq{eq:wavefunction}, which, however, is crucial to ensure the orthonormality of the new basis states~\cite{10.21468/SciPostPhys.14.6.154, JML-1-1, 10.1063/5.0209255, p3th-25bc, zhang2025quantumanharmoniceffectshydrogenbond, saleh_computing_2025}.
Compared to ground-state neural wave functions~\cite{PhysRevLett.131.126501, rbsg-r7hx, linteau2025universalneuralwavefunctions}, the Jastrow factor cannot be added to our wave functions since it would break the orthogonality.
We parametrize the two bijective maps $\vec{s} \leftrightarrow \vec{\zeta}, \vec{r} \leftrightarrow \vec{\xi}$ mentioned above using neural networks similar to the FermiNet architecture~\cite{PhysRevResearch.2.033429}. In particular, we employ permutation-equivariant layers to ensure the symmetry (anti-symmetry) property of $p(\vec{s})$ ($\Psi_{\vec{s},\vec{k}}(\vec{r})$) under permutation of nuclei (electrons with the same spin); see the Supplemental Material~\cite{supplemental} for details (see also Refs.~\cite{pyscf2018, 1997MolPh..92..477L, anie.200904748, jax2018github, 10.1063/1.464180, Zhou_2018, PhysRevResearch.2.043206, PhysRevB.111.115149, pmlr-v37-germain15, pmlr-v48-oord16, NIPS2017_3f5ee243} therein).

The electron distribution $p(\vec{k}|\vec{s})$ on the single-particle orbitals is modeled using a masked variational autoregressive network (VAN)~\cite{PhysRevLett.122.080602, 10.21468/SciPostPhys.14.6.154}. For a given nuclear configuration $\vec{s}$, the HF solver provides $M$ orthonormal single-particle orbitals $\{\psi_{\vec{s},k} \mid 1 \leq k \leq M\}$ with corresponding energy levels $\{\epsilon_{\vec{s},k} \mid 1 \leq k \leq M\}$. The occupation of these orbitals by $N/2$ spin-up and $N/2$ spin-down electrons must satisfy the Pauli exclusion principle.
As a result, the distribution $p(\vec{k}|\vec{s})$ factorizes into two components for the spin-up and spin-down sectors
\begin{align}
    p(\vec{k}|\vec{s}) = p(\vec{k}^{\uparrow}|\vec{s}) p(\vec{k}^{\downarrow}|\vec{k}^{\uparrow},\vec{s}) \, .\label{eq:transformer1}
\end{align}
We employ a mask in VAN to enforce an ordering in which electrons occupy orbitals from lowest to highest energy, i.e., $1 \leq k_1^{\sigma} < k_2^{\sigma} < \cdots < k_{N/2}^{\sigma} \leq M$. The corresponding probability is expressed as a product of conditional probabilities
\begin{subequations}
    \begin{align}
        p(\vec{k}^{\uparrow}|\vec{s}) &= \prod_{i=1}^{N/2} p(k_i^{\uparrow}|k_1^{\uparrow},\cdots,k_{i-1}^{\uparrow},\vec{s}) \, , \label{eq:transformer2a} \\
        p(\vec{k}^{\downarrow}|\vec{k}^{\uparrow},\vec{s}) &= \prod_{i=1}^{N/2} p(k_i^{\downarrow}|k_1^{\downarrow},\cdots,k_{i-1}^{\downarrow},\vec{k}^{\uparrow},\vec{s}) \, . \label{eq:transformer2b}  
    \end{align}
\end{subequations}
In order to feed the information of nuclear configuration $\vec{s}$ into the autoregressive model, we give the analytical conditional probability of non-interacting fermions on HF energy levels into the logit bias of VAN. This makes VAN start with the non-interacting electron distribution, and then learn the interaction effects via the following training. This step is an efficient alternative to the pretraining step in Ref.~\cite{10.21468/SciPostPhys.14.6.154}.
See supplemental material~\cite{supplemental} for details of the masked autoregressive model.

\begin{figure}
    \centering
    \includegraphics[width=\linewidth]{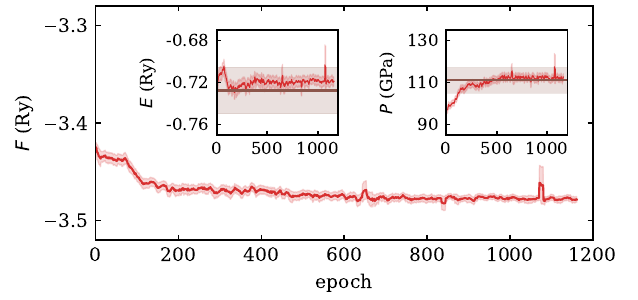}
    \caption{Training curve of variational free energy method for $N=32$, $T=31\,250$ K, $r_s=2$ deuterium system, with a 10-epoch moving average. The red line in the main figure shows the variational free energy $F$ per atom and the error bar (red-filled area) during training. The red lines in the set insets show the internal energy $E$ per atom and the pressure $P$ change. The brown lines are the PBC $N=32$ results from RPIMC~\cite{PhysRevLett.85.1890}, and the filled areas denote the error bars.}
    \label{fig:trainingcurve}
\end{figure}

The training curve of the variational free energy method is shown in \Fig{fig:trainingcurve}.
The starting point of the training curve is the pretrained flow model based on the HF potential energy surface, combined with the untrained VAN and electron backflow network. The pretraining process is discussed in the supplemental material~\cite{supplemental}.
During the training process, the three neural networks are jointly optimized to minimize the free energy defined in \Eq{eq:FreeEnergy} via stochastic optimization. At each optimization step, nucleus and electron configurations are sampled via Monte Carlo sampling from the distributions $p(\vec{s})$ and $\left|\Psi_{\vec{s},\vec{k}}(\vec{r})\right|^2$, while the energy level configurations $\vec{k}$ are drawn from $p(\vec{k}|\vec{s})$ via direct sampling.
The variational free energy is then estimated by computing the expectation values over these samples, as specified in \Eq{eq:FreeEnergy}, and the network parameters are updated via stochastic reconfiguration~\cite{Becca_Sorella_2017}.
Due to the variational principle, the approximation becomes closer to the ground truth as the free energy is minimized.
To obtain the equation of state of deuterium, we perform variational free energy optimization under different densities and temperatures. After the optimization has reached convergence, we carry out inference to evaluate various thermodynamic observables such as the entropy of nuclei and electrons, pressure, and internal energy.


\begin{table*}\scriptsize
    \centering
    \renewcommand{\arraystretch}{1.2}
    \resizebox{\textwidth}{!}{
    \begin{tabular}{ c  c  c c c   c c c  c  c  c }
        \toprule
        \multirow{2}{*}{$N$} & \multirow{2}{*}{$T$} & \multicolumn{3}{c}{$r_s=1.86$} & \multicolumn{3}{c}{$r_s=2$} & \multirow{2}{*}{$\rho_{\mathrm{Hugoniot}}/\rho_0$} & \multirow{2}{*}{$E_{\mathrm{Hugoniot}}$} & \multirow{2}{*}{$P_{\mathrm{Hugoniot}}$}  \\
        \cmidrule{3-8}
        & & $E$ & $P$ & $S_e$ & $E$ & $P$ & $S_e$ & & & \\
        \midrule
        54 & $10\,000$ & -0.9669(18) & 65.95(71) & 0.0220(4) & -0.9675(13) & 45.48(49) & 0.0271(4) & 4.495(15) & -0.9671(12) & 57.34(33) \\
        32 & $10\,000$ & -0.9726(6) & 62.85(34) & 0.0549(2) & -0.9778(6) & 43.03(28) & 0.07003(3) & 4.5225(90) & -0.9746(4) & 55.11(13) \\
        32 & $15\,625$ & -0.9050(5) & 86.27(33) & 0.1600(4) & -0.9077(17) & 60.54(51) & 0.1751(4) & 4.4730(64) & -0.9062(8) & 74.89(24) \\
        32 & $31\,250$ & -0.7105(6) & 155.12(33) & 0.4389(5) & -0.7089(5) & 115.50(26) & 0.4794(5) & 4.3481(36) & -0.7096(4) & 132.43(12) \\
        32 & $62\,500$ & -0.2939(37) & 302.0(13) & 0.8353(6) & -0.2892(7) & 233.89(28) & 0.8890(6) & 4.2438(23) & -0.2906(12) & 255.58(37) \\
        \bottomrule
    \end{tabular}
    }
    \caption{The equation of state and Hugoniot results of deuterium system calculated by variational free energy method. $N$ is the number of atoms in the system. $T$ is the temperature in Kelvin unit. $E$, $P$, and $S_e$ are the internal energy (in Rydberg unit), pressure (in GPa), and electronic entropy (in $k_{\mathrm{B}}$ unit) per atom, respectively. The subscripts denote the corresponding $r_s$ values. $\rho_{\mathrm{Hugoniot}}$, $E_{\mathrm{Hugoniot}}$, and $P_{\mathrm{Hugoniot}}$ are the Hugoniot density (in units of initial density $\rho_0 = 0.171~\text{g/cm}^3$), internal energy (in Rydberg unit), and pressure (in GPa), respectively.}
    \label{tab:eos}
\end{table*}

\section{Result}
\begin{figure}
    \centering
    \includegraphics[width=1\linewidth]{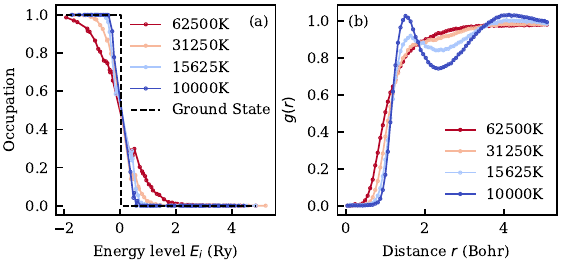}
    \caption{Electron distribution on single-particle orbitals and nucleus-nucleus RDF of $N=32$, $r_s=2$ deuterium system at several temperatures. (a) Electron occupation of single-particle orbitals. $E_i$ are Hartree-Fock energy levels of these orbitals. The dashed black line is ground state occupation as reference. (b) Nucleus-nucleus RDF.
    }
    \label{fig:occ_rdf}
\end{figure}

To study the EOS of the deuterium system, we perform calculations on systems containing $14$, $20$, $32$, and $54$ atoms. All calculations are carried out using twist boundary conditions (TBC), with the Baldereschi point $(1/4, 1/4, 1/4)$ as the mean-value point. We also apply Chiesa's finite-size correction~\cite{PhysRevLett.97.076404,PhysRevE.81.021202} to reduce finite-size errors. See supplemental material~\cite{supplemental} for details.
\Table{tab:eos} presents results for systems with $N = 32$ and $54$ atoms at various $r_s$ and temperatures.
As shown in \Table{tab:eos}, the electronic entropy is close to zero at $10\,000$ K, indicating that most electrons remain in the ground state. As the temperature increases, electrons are gradually thermally excited to higher energy levels. 
    We can also see this intuitively in \Fig{fig:occ_rdf}(a), which shows the electron occupation of single-particle orbitals. At $10\,000$ K, the electron occupation closely resembles that of the ground state (the dashed black line). As temperature rises, electrons increasingly occupy higher energy levels, leading to a broader distribution. This Fermi-Dirac-like distribution of electrons at finite temperature is captured by VAN.
The discontinuous jumps in the occupation at certain energy levels arise from the learning of the neural canonical transformation. After the transformation in \Eq{eq:wavefunction}, the notions of single-particle orbitals and energy levels are no longer well defined for the many-body wave function. Instead, the VAN determines whether to occupy these transformed orbitals based on their contribution to lowering the system's free energy, irrespective of their original Hartree-Fock energy levels.
The entropy and occupation results indicate that thermal effects on electrons must be taken into account when the temperature is above $10\,000$ K.

The nucleus-nucleus radial distribution function (RDF) $g(r)$ is shown in \Fig{fig:occ_rdf}(b). At $62\,500$ K, $g(r)$ is smooth without any prominent peak, indicating a fully dissociated atomic fluid phase. As the temperature decreases, a peak corresponding to the formation of deuterium molecules gradually emerges.
This marks the beginning of the transition from the atomic phase to the molecular phase~\cite{PhysRevB.110.L041107}.
The formation of the molecular peak in RDF can be verified by previous variational density matrix calculation~\cite{PhysRevE.61.3470}.

\begin{figure}
    \centering
    \includegraphics[width=0.85\linewidth]{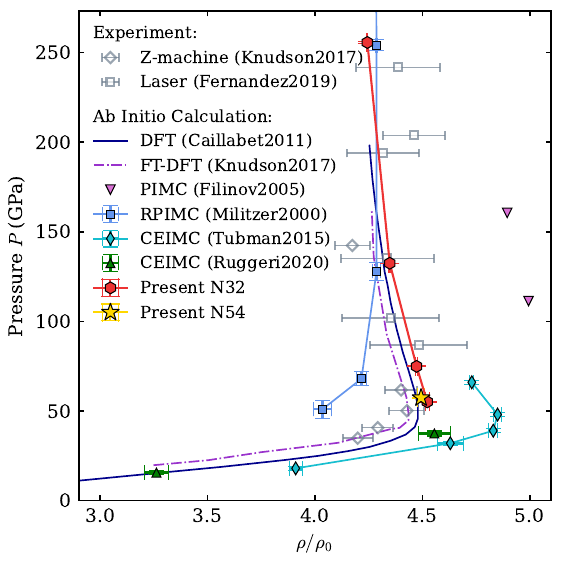}
    \caption{Pressure-compression ratio diagram of Hugoniot experiments and computations. The red symbols are Lagrange interpolation of variational free energy results of $N=32$ system, the yellow star is the result at $10\,000$ K for the $N=54$ system. The experimental and theoretical results are from Z-machine~\cite{PhysRevLett.118.035501}, laser~\cite{PhysRevLett.122.255702}, DFT molecular dynamics (DFT)~\cite{PhysRevB.83.094101}, DFT with finite temperature XC functionals (FT-DFT)~\cite{PhysRevLett.118.035501}, direct PIMC~\cite{filinov_calculation_2005}, RPIMC~\cite{PhysRevLett.85.1890}, and CEIMC~\cite{PhysRevLett.115.045301,PhysRevB.102.144108}.
    }
    \label{fig:hugoniot}
\end{figure}

We compute the deuterium Hugoniot curve using the equation of state and the Rankine-Hugoniot relation~\cite{RevModPhys.49.523}
\begin{equation}
    H = E-E_0+\frac{1}{2}(P+P_0)(\Omega-\Omega_0)=0 \, .\label{eq:Hugoniot}
\end{equation}
The reference state used in this work is $\rho_0=0.171~\text{g/cm}^3$, $E_0 = -15.886~\text{eV}$, and $P_0 = 0~\text{GPa}$, which is the same as that adopted in Ref.~\cite{PhysRevLett.85.1890}. 
The deuterium Hugoniot curve is typically located in the range $r_s = 1.86$ to $r_s = 2$ for temperatures between $10\,000$ K and $62\,500$ K. We select four representative temperatures for the calculations. At each temperature, the EOS points are computed at $r_s = 1.86$ and $r_s = 2$.
The Hugoniot density and corresponding pressure are then obtained via Lagrange interpolation, as summarized in \Table{tab:eos}.
The interpolated Hugoniot curve for the $N = 32$ and $N = 54$ deuterium system is shown in \Fig{fig:hugoniot}. The yellow star denotes the result at $10\,000$ K for the $N = 54$ system, which corresponds to the regime where finite-size effects are expected to be most pronounced, as shell effects are particularly strong at low temperatures. This computationally expensive calculation is included to benchmark the magnitude of finite-size effects in our study.
Our results are consistent with the available experimental data~\cite{PhysRevLett.118.035501, PhysRevLett.122.255702}. As shown in \Fig{fig:hugoniot}, at low temperatures, our Hugoniot compression ratio $\rho/\rho_0$ is slightly higher than the DFT results, significantly higher than those from RPIMC~\cite{PhysRevLett.85.1890}, and substantially lower than the CEIMC results~\cite{PhysRevLett.115.045301}. With increasing temperature, our results show good agreement with RPIMC.
It should be noted that the PIMC method suffers from the fermion sign problem~\cite{PhysRevE.100.023307,PhysRevE.107.015206} at low temperatures and small $r_s$, where electron-electron interactions become strong. Moreover, the RPIMC method relies on the fixed-node approximation, which can introduce systematic errors and requires additional effort to determine appropriate nodal surfaces. In contrast, the variational free energy method employed here is free from the fermion sign problem and does not impose any restricted approximations on the electronic degrees of freedom. The computational procedure remains the same across different temperatures and values of $r_s$.

\section{Conclusion}
In this work, we further develop the variational free energy method using three generative neural networks to parameterize the density matrix of the warm dense deuterium system. By optimizing the variational free energy with respect to three neural networks, we calculate the EOS and Hugoniot curve of deuterium within the temperature range where electrons are slightly excited ($10\,000\sim62\,500$ K). We focus on this temperature regime to bridge the gap from the finite-temperature side, as existing ground-state methods become unreliable when thermal electron excitations are significant, while PIMC methods suffer from the fermion sign problem at lower temperatures.
The advantage of the variational free energy method compared to PIMC is that there is no exponential computational complexity caused by the fermion sign problem, and it can directly compute entropy and free energy. 
Compared to CEIMC, the variational free energy method effectively accounts for finite-temperature effects on electrons, which is critical for studying WDM.
Moreover, the variational principle ensures its reliability. 
The resulting Hugoniot curve, computed with twisted boundary conditions and analytical finite-size corrections for systems up to 54 atoms, shows good agreement with experimental measurements, connects smoothly with PIMC results at high temperatures, and extends into the low-temperature regime where PIMC suffers from the fermion sign problem, thereby achieving the handshake between ground-state and finite-temperature calculations from the finite-temperature side.
In conclusion, our results provide the EOS and Hugoniot curve of deuterium for benchmarking future experiments and theoretical methods. We note that recent advances in PIMC methods~\cite{10.1063/5.0106067, morresi2025study, 10.1063/5.0171930, dornheim2025fermionicfreeenergiestextitab, dornheim2024ab, PhysRevB.111.L041114, 4n7x-78fs, svensson2025acceleratedfreeenergyestimation} may further verify our results.

\begin{acknowledgments}
We are grateful for the useful discussions with Xinguo Ren, Mohan Chen, Garnet Chan, Shiwei Zhang, Markus Holzmann, Michele Casula, Qi Zhang, Zhendong Cao, Xingyu Zhang, Qi Yang, and Ruisi Wang. This work is supported by the National Natural Science Foundation of China under Grants No. 92270107, No. T2225018, No. 12188101, No. T2121001,  the Strategic Priority Research Program of the Chinese Academy of Sciences under Grants No. XDB0500000, and the National Key Projects for Research and Development of China Grants No. 2021YFA1400400. 
\end{acknowledgments}

\bibliographystyle{apsrev4-2}
\bibliography{refs}
\end{document}


\title{Supplemental Material: Deep Variational Free Energy Calculation of Hydrogen Hugoniot}

\author{Zihang Li}
\affiliation{Beijing National Laboratory for Condensed Matter Physics and Institute of Physics, \\Chinese Academy of Sciences, Beijing 100190, China}
\affiliation{University of Chinese Academy of Sciences, Beijing 100049, China}

\author{Hao Xie}
\email{qwexiehao@gmail.com}
\affiliation{Beijing National Laboratory for Condensed Matter Physics and Institute of Physics, \\Chinese Academy of Sciences, Beijing 100190, China}
\affiliation{Department of Astrophysics, University of Zürich, Winterthurerstrasse 190, 8057 Zürich, Switzerland}

\author{Xinyang Dong}
\email{dongxy@iphy.ac.cn}
\affiliation{Beijing National Laboratory for Condensed Matter Physics and Institute of Physics, \\Chinese Academy of Sciences, Beijing 100190, China}

\author{Lei Wang}
\email{wanglei@iphy.ac.cn}
\affiliation{Beijing National Laboratory for Condensed Matter Physics and Institute of Physics, \\Chinese Academy of Sciences, Beijing 100190, China}

\maketitle
\setcounter{table}{0}
\renewcommand{\thetable}{S\arabic{table}}
\setcounter{figure}{0}
\renewcommand{\thefigure}{S\arabic{figure}}
\setcounter{equation}{0}
\renewcommand{\theequation}{S\arabic{equation}}

\setcounter{section}{0}
\renewcommand{\thesection}{S\arabic{section}}
\setcounter{secnumdepth}{3}


\section{Derivation of variational free energy}
\label{sec:derivation}
The variational free energy is the loss function of the variational free energy method. The derivation of \Eq{eq:FreeEnergy} starts with the density matrix of the system
\begin{equation}
    \rho = \int d{\vec{s}}\sum_{\vec{k}} p(\vec{s},\vec{k})\ket{\vec{s},\Psi_{\vec{s},\vec{k}}}\bra{\vec{s},\Psi_{\vec{s},\vec{k}}} \, .\label{eq:DensityMatrix2}
\end{equation}
The density matrix $\rho$ is expressed as a sum over the probability distribution $p(\vec{s},\vec{k})$ multiplied by the corresponding many-body state $\ket{\vec{s},\Psi_{\vec{s},\vec{k}}}$. The joint probability $p(\vec{s},\vec{k})$ is given by $p(\vec{s},\vec{k}) = p(\vec{s}) p(\vec{k}|\vec{s})$, where $p(\vec{s})$ is the normalized Boltzmann distribution of nuclei, satisfying $\int_{\vec{s}} p(\vec{s})d\vec{s} = 1$, and $p(\vec{k}|\vec{s})$ is the normalized conditional probability of electron many-body states for a given nuclear configuration $\vec{s}$, satisfying $\sum_{\vec{k}} p(\vec{k}|\vec{s}) = 1$. Thus, we get \Eq{eq:DensityMatrix}. By substituting the density matrix into the free energy expression, the variational free energy is obtained as
\begin{align}
    F\left[\rho\right] &= \frac{1}{\beta}\Tr\left(\rho\ln\rho\right) + \Tr\left(\rho H\right)\notag\\
    &= \int d{\vec{s}}\sum_{\vec{k}} p(\vec{s},\vec{k}) \left[\frac{1}{\beta}\ln p(\vec{s},\vec{k}) + \bra{\vec{s},\Psi_{\vec{s},\vec{k}}}H\ket{\vec{s},\Psi_{\vec{s},\vec{k}}}\right]\notag\\
    &= \int d{\vec{s}}\ p(\vec{s})\sum_{\vec{k}} p(\vec{k}|\vec{s}) \left[\frac{1}{\beta}\ln p(\vec{s}) + \frac{1}{\beta}\ln p(\vec{k}|\vec{s}) + E(\vec{s},\vec{k})\right]\notag\\
    &= \int d{\vec{s}}\ p(\vec{s}) \left\{\frac{1}{\beta}\ln p(\vec{s}) + \sum_{\vec{k}}p(\vec{k}|\vec{s})\left[\frac{1}{\beta}\ln p(\vec{k}|\vec{s}) + E(\vec{s},\vec{k})\right]\right\} \, .\label{eq:FreeEnergy2}
\end{align}
Here, $E(\vec{s},\vec{k})$ represents the energy, defined as the expectation value of the Hamiltonian $H$ for the state $\ket{\vec{s},\Psi_{\vec{s},\vec{k}}}$, which is expressed as
\begin{equation}
    E(\vec{s},\vec{k}) = \int d\vec{r}\left|\Psi_{\vec{s},\vec{k}}(\vec{r})\right|^2\frac{\mathcal{H}\Psi_{\vec{s},\vec{k}}(\vec{r})}{\Psi_{\vec{s},\vec{k}}(\vec{r})} \, .\label{eq:Energy}
\end{equation}
The variational free energy in \Eq{eq:FreeEnergy2} is expressed as the sum of three components
\begin{equation}
    F = \int d\vec{s}\ p(\vec{s}) \left\{\frac{1}{\beta}\ln p(\vec{s}) + \sum_{\vec{k}}p(\vec{k}|\vec{s})\left[\frac{1}{\beta}\ln p(\vec{k}|\vec{s}) + \int d\vec{r}\left|\Psi_{\vec{s},\vec{k}}(\vec{r})\right|^2\frac{\mathcal{H}\Psi_{\vec{s},\vec{k}}(\vec{r})}{\Psi_{\vec{s},\vec{k}}(\vec{r})}\right]\right\} \, .\label{eq:FreeEnergy3}
\end{equation}
The three summations in \Eq{eq:FreeEnergy3} can be expressed as three expectation values, yielding the variational free energy in \Eq{eq:FreeEnergy}.

\section{Batched Hartree-Fock solver}
\label{sec:hf}
In the variational free energy method, we need some orthogonal single-particle orbitals to construct Slater determinants in expressing the electron wave function. The electron wave function takes orthogonal single-particle orbitals as basis functions. 
The orthogonality of the orbitals is a necessary requirement, as it ensures the physical consistency of the electronic entropy and the validity of the variational principle. 
Based on the premise of satisfying orthogonality, the criterion for judging the quality of a orbital is whether it can lead to a lower free energy.
For periodic structures, we have tried common orthogonal electron orbitals include plane wave orbitals, HF orbitals solved with and without exchange, DFT orbitals solved with different XC functionals, etc.
The results show that HF orbitals can lead to the lowest free energy among these choices.

To obtain the Hartree-Fock orbitals, we need to solve the Hartree-Fock equation while the nuclei are moving. The HF solution is computationally expensive, especially for large systems. To address this issue, we develop a batched Hartree-Fock solver that can compute the HF orbitals and energy levels for a batch of nuclear configurations in parallel. This significantly speeds up the calculation, making it feasible to perform variational free energy calculations for warm dense matter systems. In addition, the batched HF solver can also be used to generate the HF potential energy surface (PES), which is used to pretrain the nucleus flow model. See \ref{sec:training} for details.

Our HF solver, similar to PYSCF~\cite{pyscf2018}, employs Gaussian-type orbitals (GTOs) as the basis to compute the HF linear combination of atomic orbitals (LCAO) and energy levels. For a batch of nuclear configurations $\vec{s}$, it calculates the HF orbitals $\psi_{\vec{s},k}$ and energy levels $\epsilon_{\vec{s},k}$ in seconds. These orbitals are used in the electron wave function in \Eq{eq:wavefunction}, while the energy levels are input to VAN. Calculations can be performed both under periodic boundary conditions (PBC) and twist boundary conditions (TBC). 

We utilize the GTH-DZV basis set, a double-zeta valence basis, for HF calculations. The choice of basis has been carefully studied. During the method development phase, we tested various GTOs including STO-3G, GTH-SZV, GTH-DZV, GTH-TZV, GTH-DZVP, and more. The conclusion is that single-zeta basis sets like STO-3G and GTH-SZV, which provide orbitals with twice the number of electrons, are not sufficient to reach converged results. Training networks on HF orbitals solved with double-zeta basis like GTH-DZV, which provide orbitals with four times the number of electrons, can reach lower free energy compared to single-zeta basis sets. Further increasing the basis set size to triple-zeta (GTH-TZV) or adding polarization functions (GTH-DZVP) does not improve the results, indicating that double-zeta basis sets are sufficient for our purpose. We have also tested the shape of GTH-DZV's Gaussians, finding that slightly changing the exponents of the Gaussians does not affect the final results. This indicates that the basis set is complete for our purpose. We believe we have achieved solid results in the relevant temperature range between $10\,000$ K and $62\,500$ K. 

The HF solver in variational free energy method are used after sampling the nuclear configuration $\vec{s}$. The nuclear coordinates are input into the HF solver, which outputs the coefficient matrix $\mathbf{C}$ and energy levels $\epsilon_{\textbf{s},k}$. The matrix elements $\mathrm{C}_{\mu k}$ represent the linear combination coefficients of the atomic orbitals $\phi_{\mu}(\vec{r})$. The HF orbital $\psi_{k}(\vec{r})$ is given by
\begin{equation}
    \psi_{k}(\vec{r}) = \sum_{\mu}\mathrm{C}_{\mu k}\phi_{\mu}(\vec{r}) \, .
\end{equation}
The coefficient matrix $\mathbf{C}$ is obtained by solving the HF Roothaan equation
\begin{equation}
    \mathbf{FC}=\mathbf{SC}\epsilon \, ,\label{eq:Roothaan}
\end{equation}
where $\mathbf{F}$ is the Fock matrix, $\mathbf{S}$ is the overlap matrix, and $\epsilon$ is the eigenvalue vector. The Fock matrix $\mathbf{F}$ is expressed as
\begin{equation}
    \mathbf{F} = \mathbf{T} + \mathbf{V}_{\mathrm{nuc}} + \mathbf{J} - \frac{1}{2}\mathbf{K} \, .
\end{equation}
Here, $\mathbf{T}$ is the kinetic matrix, $\mathbf{V}_{\mathrm{nuc}}$ is the nuclear potential matrix, $\mathbf{J}$ and $\mathbf{K}$ are the Hartree and exchange matrices, respectively. The overlap matrix $\mathbf{S}$ and kinetic matrix $\mathbf{T}$ are computed efficiently using the Gaussian product theorem.

The most computationally intensive part of the HF calculation is the four-center two-electron repulsion integrals (ERIs) in the exchange term. We employ the Gaussian Plane Waves (GPW) method~\cite{1997MolPh..92..477L, anie.200904748} to compute these integrals for GTOs. The integrals are defined as
\begin{equation}
    \left(\mu\nu|\kappa\lambda\right) = \int_{\Omega}d\vec{r}_1\int d\vec{r}_2\ \phi_{\mu}^*(\vec{r}_1)\phi_{\nu}(\vec{r}_1)g(\vec{r}_{12})\phi_{\kappa}^*(\vec{r}_2)\phi_{\lambda}(\vec{r}_2) \, .
\end{equation}
The Hartree and exchange terms are given by
\begin{align}
    &\mathrm{J}_{\mu\nu} = \sum_{\kappa\lambda}\mathrm{P}_{\kappa\lambda}(\mu\nu|\kappa\lambda) \, ,\\
    &\mathrm{K}_{\mu\nu} = \sum_{\kappa\lambda}\mathrm{P}_{\kappa\lambda}(\mu\kappa|\lambda\nu) \, ,
\end{align}
where $\mathrm{P}_{\kappa\lambda} = 2 \sum_{k}^{\mathrm{occ}} \mathrm{C}_{\kappa k} \mathrm{C}_{\lambda k}^*$ is the density matrix. The electron repulsion integrals are computed using GPW method, expressed as
\begin{equation}
    \left(\mu\nu|\kappa\lambda\right) = \frac{1}{\Omega}\sum_{\textbf{G}}\mathscr{C}(\textbf{G})\rho_{\mu\nu}(\textbf{G})\rho_{\kappa\lambda}(-\textbf{G}) \, .\label{eq:eri}
\end{equation}
The Fourier-transformed Coulomb kernel is
\begin{equation}
    \mathscr{C}(\textbf{G}) = \left\{
    \begin{aligned}
        &\frac{4\pi}{|\textbf{G}|^2}\quad\text{if}\quad |\textbf{G}| \neq 0\\
        &0\qquad\ \ \text{if}\quad |\textbf{G}| = 0
    \end{aligned}
    \right. \, .
\end{equation}
The Fourier-transformed orbital product $\rho_{\mu\nu}(\textbf{G})$ is
\begin{equation}
    \rho_{\mu\nu}(\textbf{G}) = \int_{\Omega}d\vec{r}\ \phi_{\mu}(\vec{r})\phi_{\nu}^*(\vec{r})e^{-i\textbf{G}\cdot\vec{r}} \, .
\end{equation}
The nuclear Coulomb potential matrix can also be given by the Fourier-transformed orbital product
\begin{equation}
    \mathrm{V}_{\mu\nu} = \frac{1}{\Omega}\sum_{\textbf{G}}\mathscr{C}(\textbf{G})S(\textbf{G})\rho_{\nu\mu}(-\textbf{G}) \, ,\label{eq:potential}
\end{equation}
where $S(\mathbf{G}) = \sum_{I} e^{-i \mathbf{G} \cdot \vec{s}_I}$ is the structure factor. The orbital product $\rho_{\mu\nu}(\mathbf{G})$ is calculated using the Fast Fourier Transform (FFT) method. 
Typically, FFT uses a grid density of 0.12 Bohr (adopted by PYSCF~\cite{pyscf2018}) in real space to ensure high-precision calculation of ERIs. For example, a system with 54 hydrogen atoms at $r_s=2$ is in a cubic box of length 
12.2 Bohr, requiring a $(12.2/0.12)^3 \approx 101^3$ FFT grid to achieve the desired grid density.
To optimize performance and reduce memory usage, we adopt a coarser FFT grid length of 0.5 Bohr, significantly lowering memory requirements. By using this grid length, the system with 54 hydrogen atoms at $r_s=2$ only requires a $(12.2/0.5)^3 \approx 25^3$ FFT grid.
This reduces GPU RAM cost to approximately $(0.12/0.5)^3 \approx 0.014$ times the original, enabling batch HF calculations for up to 54 hydrogen atoms.
This parameter change is equivalent to ignoring the high-frequency terms in the summation of \Eq{eq:eri} and \Eq{eq:potential}, which affects the Coulomb interaction components $\mathbf{V}_{\mathrm{nuc}}$, $\mathbf{J}$, and $\mathbf{K}$ in Fock matrix $\mathbf{F}$.

To quantify the impact, we conducted a controlled experiment comparing two grid lengths (0.2 Bohr and 0.5 Bohr) for the HF solver during network training and inference, while keeping all other parameters constant. The impact on the final equation of state at $N=32$, $T=10\,000$ K and $r_s=1.86$ under PBC is shown in the \Table{tab:grid_comparison}.
The results show that the differences in free energy and other thermodynamic quantities are small and within statistical uncertainty. Remarkably, the coarser grid (0.5 Bohr) even yields a slightly lower free energy, which is the variational criterion for quality.

\begin{table}[H]
    \centering
    \renewcommand{\arraystretch}{1.2}
    \resizebox{\textwidth}{!}{
    \begin{tabular}{ c c c c c c c c c c }
        \hline
        $l_{\mathrm{grid}}$ & $F\ (\mathrm{Ry})$ & $E\ (\mathrm{Ry})$ & $V_{en}\ (\mathrm{Ry})$ & $V_{ee}\ (\mathrm{Ry})$ & $V_{nn}\ (\mathrm{Ry})$ & $K_e\ (\mathrm{Ry})$ & $P\ (\mathrm{GPa})$ & $S_e\ (k_{\mathrm{B}})$ & $S_n\ (k_{\mathrm{B}})$ \\
        \hline
        0.2 & -1.77839(83) & -0.98925(83) & -1.1815(17) & -0.46957(55) & -0.64055(24) & 1.2073(19) & 56.96(43) & 0.04276(23) & 12.41674(38) \\
        0.5 & -1.77966(61) & -0.98913(61) & -1.1735(18) & -0.47085(54) & -0.63946(24) & 1.1997(17) & 55.59(35) & 0.06456(24) & 12.41704(39) \\
        \hline
    \end{tabular}
    }
    \caption{Comparison of variational free energy and thermodynamic quantities obtained using different FFT grid lengths in the HF solver at $N=32$, $T=10\,000$ K and $r_s=1.86$ under PBC.}
    \label{tab:grid_comparison}
\end{table}

Importantly, the change of GPW FFT grid length preserves the orthogonality of the HF orbitals, which depends solely on the accuracy of the overlap matrix $\mathbf{S}$ of Gaussian-type orbitals (this will be proved in \ref{sec:orthonormality}), not on the FFT grid length. We compute $\mathbf{S}$ using a sufficiently large cutoff radius to ensure its accuracy. For the temperature range considered ($T = 10\,000$--$62\,500$ K), the particle distribution is relatively uniform, which means that ignoring the high-frequency terms in reciprocal space does not lead to significant loss of accuracy in the HF orbitals. Furthermore, the flow model for electron wave functions learns features directly from nuclear and electronic coordinates, allowing it to compensate for small inaccuracies in the HF orbitals. Controlled experiments confirm that the change of GPW FFT grid length has negligible impact on the final equation of state and other thermodynamic quantities.

The HF solver employs Fermi-Dirac smearing and the DIIS method to improve convergence of the self-consistent equation, achieving convergence for most configurations. Fully implemented in JAX~\cite{jax2018github}, it enables efficient GPU parallelization. 
The HF solver is open source and can be found at Ref.~\cite{hqc}.

\section{Orthonormality of wave function}
\label{sec:orthonormality}
The orthogonality of electron wave functions $\Psi_{\vec{s},\vec{k}}(\vec{r})$ is essential for the validity of the variational free energy method.
If orthogonality is broken, the free energy can reach a value lower than that of the natural world, because the variational principle is broken and the next variational lower bound no longer applies.
Besides, it is also essential to ensure easy access to the electronic entropy. 
Two key factors ensure the orthonormality of electron wave functions: the orthonormality of single-particle orbitals, and the Jacobian factor involving the square root in \Eq{eq:wavefunction}. 
Regarding the former, as we have claimed in the \ref{sec:hf}, the orthogonality of HF orbitals is guaranteed by the accuracy of the overlap matrix $\mathbf{S}$, regardless of convergence or grid length. Here is the proof.

The overlap matrix element of the HF orbitals is given by
\begin{equation}
    \Delta_{kk'} = \int_{\Omega}\ \psi_{k}^*(\vec{r})\psi_{k'}(\vec{r})d\vec{r} = \sum_{\mu,\nu}\mathrm{C}_{\mu k}^*\mathrm{C}_{\nu k'}\mathrm{S}_{\mu\nu} \, ,
\end{equation}
or in matrix form
\begin{equation}
    \mathbf{\Delta} = \textbf{C}^{\dagger}\textbf{SC} \, .
\end{equation}
To solve the generalized eigenvalue problem in \Eq{eq:Roothaan}, the first step is to orthogonalize overlap matrix $\mathbf{S}$
\begin{equation}
    \textbf{SU} = \textbf{Us} \, ,
\end{equation}
where $\mathbf{U}$ is the unitary eigenvector matrix of $\mathbf{S}$ and $\mathbf{s}$ is the diagonal eigenvalue matrix. Define matrix $\mathbf{X} \equiv \mathbf{Us}^{-1/2}$, then the orthogonalized Fock matrix $\mathbf{F'}$ and coefficient matrix $\mathbf{C'}$ given by
\begin{equation}
    \mathbf{F'} = \mathbf{X}^{\dagger}\mathbf{FX} \, .
\end{equation}
\begin{equation}
    \mathbf{C'} = \mathbf{X^{-1}C} \, .
\end{equation}
The Roothaan equation in \Eq{eq:Roothaan} is transformed into a standard eigenvalue problem
\begin{equation}
    \mathbf{F'C'} = \mathbf{C'}\epsilon \, ,
\end{equation}
where $\mathbf{C'}$ is an unitary matrix.
So the overlap matrix of HF orbitals can be expressed as
\begin{equation}
    \mathbf{\Delta} = \mathbf{C}^{\dagger}\mathbf{SC} = \mathbf{C'}^{\dagger}\mathbf{X}^{\dagger}\mathbf{SX}\mathbf{C'} = \mathbf{C'}^{\dagger}\mathbf{C'} = \mathbf{I} \, .\label{eq:orthonormality}
\end{equation}
The equalities in \Eq{eq:orthonormality} hold if $\mathbf{C}$ satisfies the Roothaan equation \Eq{eq:Roothaan} and overlap matrix $\mathbf{S}$ is accurately calculated, regardless of SCF convergence or FFT grid length.

With orthonormal HF orbitals, the orthonormality of $\Psi_{\vec{s},\vec{k}}(\vec{r})$ is established through the following derivation
\begin{align}
    \int d\vec{r}\Psi_{\vec{s},\vec{k}}^*(\vec{r})\Psi_{\vec{s},\vec{k}'}(\vec{r}) &= \int d\vec{r}\det\left[\psi_{\vec{s},k_i^{\uparrow}}^*(\boldsymbol{\xi}_j^{\uparrow})\right]\det\left[\psi_{\vec{s},k_i^{\downarrow}}^*(\boldsymbol{\xi}_j^{\downarrow})\right]\det\left[\psi_{\vec{s},k_i^{'\uparrow}}(\boldsymbol{\xi}_j^{\uparrow})\right]\det\left[\psi_{\vec{s},k_i^{'\downarrow}}(\boldsymbol{\xi}_j^{\downarrow})\right]\left|\det\left(\frac{\partial\boldsymbol{\xi}}{\partial \vec{r}}\right)\right|\notag\\
    &= \left\{\int d\boldsymbol{\xi}^{\uparrow}\det\left[\psi_{\vec{s},k_i^{\uparrow}}^*(\boldsymbol{\xi}_j^{\uparrow})\right]\det\left[\psi_{\vec{s},k_i^{'\uparrow}}(\boldsymbol{\xi}_j^{\uparrow})\right]\right\}\left\{\downarrow\right\}\notag\\
    &= \left\{\int d\boldsymbol{\xi}^{\uparrow}\left[\sum_{\tau\in S_{n/2}}\mathrm{sgn}(\tau)\ \prod_{i=1}^{n/2} \psi_{\vec{s},k_{\tau(i)}^{\uparrow}}^*(\boldsymbol{\xi}_i^{\uparrow})\right]\left[\sum_{\iota\in S_{n/2}}\mathrm{sgn}(\iota)\ \prod_{j=1}^{n/2} \psi_{\vec{s},k_{\iota(j)}^{'\uparrow}}(\boldsymbol{\xi}_j^{\uparrow})\right]\right\}\left\{\downarrow\right\}\tag{*}\label{eq:Leibniz}\\
    &= \left\{\sum_{\tau,\iota\in S_{n/2}}\mathrm{sgn}(\tau)\ \mathrm{sgn}(\iota)\ \prod_{i=1}^{n/2}\int d\boldsymbol{\xi}_i^{\uparrow}\psi_{\vec{s},k_{\tau(i)}^{\uparrow}}^*(\boldsymbol{\xi}_i^{\uparrow})\psi_{\vec{s},k_{\iota(i)}^{'\uparrow}}(\boldsymbol{\xi}_i^{\uparrow})\right\}\left\{\downarrow\right\}\notag\\
    &= \left\{\sum_{\tau,\iota\in S_{n/2}}\mathrm{sgn}(\tau)\ \mathrm{sgn}(\iota)\ \prod_{i=1}^{n/2}\delta_{k_{\tau(i)}^{\uparrow}k_{\iota(i)}^{'\uparrow}}\right\}\left\{\downarrow\right\}\notag\\
    &= \left\{\sum_{\tau,\iota\in S_{n/2}}\mathrm{sgn}(\tau)\ \mathrm{sgn}(\iota)\ \delta_{k_{\tau}^{\uparrow}k_{\iota}^{'\uparrow}}\right\}\left\{\downarrow\right\}\notag\\
    &= \delta_{\vec{k}^{\uparrow}\vec{k}^{'\uparrow}}\delta_{\vec{k}^{\downarrow}\vec{k}^{'\downarrow}}\notag\\
    &= \delta_{\vec{k}\vec{k}'} \, .\label{eq:orthonormality1}
\end{align}
In the derivation (\ref{eq:Leibniz}), we apply the Leibniz formula for Slater determinants.

\section{Conditional probability of ideal fermion}
\label{sec:ideal_fermion}
The conditional probability of ideal fermion is used to give the HF electron distribution on HF orbitals to VAN. This can be derived from the Fermi-Dirac statistics of a system comprising non-interacting, identical fermions, analyzed within the canonical ensemble.
We use dynamic programming in the derivation. Similar discussions of the partition function of ideal fermion gas are in Refs.~\cite{10.1063/1.464180, Zhou_2018, PhysRevResearch.2.043206, PhysRevB.111.115149}. The difference between this derivation and others is discussed at the end of this section.

Consider a many-particle system of $N$ identical non-interacting fermions in thermal equilibrium. Spin is ignored in this section, or we can assume all the fermions have the same spin. The system has $M$ single-particle states, indexed by $k = 1, 2, \cdots, M$, where the $k$-th state has energy $\epsilon_k$. The system must satisfy $N<M$, or the excess fermions will have no states to occupy.
We arrange these energy levels $\epsilon_k$ in ascending order in the variational free energy method, but in this section, this is not necessary.
The occupied states are denoted by the index set $\vec{k} = \{k_1, k_2, \cdots, k_{N}\}$. The system has $C_M^{N}$ possible many-body states. To simplify the problem, we enforce the ordering $1 \leq k_1 < k_2 < \cdots < k_{N} \leq M$. The total energy of the state $\vec{k}$ is given by
\begin{equation}
    E_{\vec{k}} = \epsilon_{k_1}+\epsilon_{k_2}+\cdots+\epsilon_{k_{N}} \, .
\end{equation}
The partition function for the system is
\begin{equation}
    \mathcal{Z} = \sum_{\vec{k}}e^{E_{\vec{k}}} \, ,
\end{equation}
where the summation extends over all possible states $\vec{k}$ of the many-particle system. The probability that the system occupies the state $\vec{k}$ is
\begin{equation}
    p_f^N(\vec{k}) = \frac{e^{-\beta E_{\vec{k}}}}{\mathcal{Z}} \, .
\end{equation}
We use $p_f^N$ to represent the probability of $N$ ideal fermions. This joint probability can be expressed as a product of conditional probabilities
\begin{equation}
    p_f^N(\vec{k}) = p_f^N(k_1) p_f^N(k_2|k_1) p_f(k_3|k_1,k_2)\cdots p_f^N(k_{N}|k_1,k_2,\cdots,k_{N-1}) \, ,\label{eq:conditional}
\end{equation}
where the first term, $p_f^N(k_1)$, represents the probability distribution of $k_1$, the lowest index of occupied state. The subsequent terms, such as $p_f^N(k_2|k_1)$, denote the conditional probability of occupying state $k_2$ given $k_1$, and so forth. We define the partition function for $n$ electrons occupying the last $m$ energy levels (from $M-m+1$ to $M$) as $\mathcal{Z}_m(n)$. Note that $m$ and $n$ satisfy $m<M$, $n<N$, and $n<m$. It can be shown that
\begin{equation}
    p_f^N(k_1) = \frac{e^{-\beta\epsilon_{k_1}}\mathcal{Z}_{M-k_1}(N-1)}{\mathcal{Z}} \, .\label{eq:qk1}
\end{equation}
The conditional probability $p_f^N(k_i|k_1,\cdots,k_{i-1})$ is given by
\begin{equation}
    p_f^N(k_i|k_1,\cdots,k_{i-1}) = \frac{e^{-\beta\epsilon_{k_i}}\mathcal{Z}_{M-k_i}(N-i)}{\mathcal{Z}_{M-k_{i-1}}(N-i+1)} \, .\label{eq:qki}
\end{equation}
Notably, the conditional probability $p_f^N(k_i|k_1,\cdots,k_{i-1})$ depends only on $k_{i-1}$, independent of ${k_1,\cdots,k_{i-2}}$, indicating that the sequence forms a Markov chain. We can abbreviate $p_f^N(k_i|k_1,\cdots,k_{i-1})$ as $p_f^N(k_i|k_{i-1})$ in the following discussion.
By using \Eq{eq:qk1} and \Eq{eq:qki}, the problem of calculating conditional probability turns into the problem of calculating the partition function $\mathcal{Z}_m(n)$.

The partition function for non-interacting fermions can be computed using a recursive formula. We require the partition functions $\mathcal{Z}_m(n)$ for indices within the range specified by
\begin{equation}
\begin{matrix}
\mathcal{Z}_M\left(N\right) & \mathcal{Z}_{M-1}\left(N\right) & \cdots & \cdots & \cdots & \mathcal{Z}_{N}\left(N\right) & & & \\
  & \mathcal{Z}_{M-1}\left(N-1\right) & \cdots & \cdots & \cdots & \mathcal{Z}_{N}\left(N-1\right) & \mathcal{Z}_{N-1}\left(N-1\right)& & \\
  & & \cdots & \cdots & \cdots & \cdots & \cdots & \cdots & \\
  & & & \mathcal{Z}_{M-N+1}(1) & \cdots & \mathcal{Z}_{N}(1) & \mathcal{Z}_{N-1}(1) & \cdots & \mathcal{Z}_{1}(1)
\end{matrix} \, .\label{eq:Zs}
\end{equation}
The partition function $\mathcal{Z}_{m}(n)$ can be decomposed into two contributions, depending on whether the energy level $M-m+1$ is occupied
\begin{equation}
    \mathcal{Z}_m(n) = e^{-\beta\epsilon_{M-m+1}}\mathcal{Z}_{m-1}(n-1)+\mathcal{Z}_{m-1}(n) \, .\label{eq:recursion1}
\end{equation}
This equation is mentioned in Ref.~\cite{PhysRevResearch.2.043206}. Applying \Eq{eq:recursion1} recursively yields a single-loop recursion formula
\begin{align}
    \mathcal{Z}_m(n) &= e^{-\beta\epsilon_{M-m+1}}\mathcal{Z}_{m-1}(n-1)+\mathcal{Z}_{m-1}(n)\notag\\
    &= e^{-\beta\epsilon_{M-m+1}}\mathcal{Z}_{m-1}(n-1)+e^{-\beta\epsilon_{M-m+2}}\mathcal{Z}_{m-2}(n-1)+\mathcal{Z}_{m-2}(n)\notag\\
    &=\cdots\notag\\
    &= \sum_{k=1}^{m-n+1}e^{-\beta\epsilon_{M-m+k}}\mathcal{Z}_{m-k}(n-1) \, .\label{eq:recursion2}
\end{align}
\Eq{eq:recursion2} means each element in \Eq{eq:Zs} can be given by a summation over elements on the right side in the next line. This is a new recursion formula that differs from other methods since we fix the order of $\vec{k}$ in the derivation. Using \Eq{eq:recursion2}, the recursion eventually reaches the terms in the bottom row of the partition function table in \Eq{eq:Zs}, which are straightforward
\begin{equation}
    \mathcal{Z}_m(1) = \sum_{k=M-m+1}^{M}e^{-\beta\epsilon_k} \, .
\end{equation}
By applying \Eq{eq:recursion2} recursively, we can compute all partition functions in \Eq{eq:Zs}.

Compared with other partition function calculation formulas, \Eq{eq:recursion2} is the most suitable one for our method. The recursion formula in Ref.~\cite{10.1063/1.464180} has a sign alternation, which can lead to sign problem when the number of particles is large. Besides, it gets the partition function recursively in the column direction, and some of the intermediate partition functions are not what we need. The formula in Refs.~\cite{Zhou_2018, PhysRevB.111.115149} is good for computing one partition function, but to get the whole table of \Eq{eq:Zs}, it has a large amount of repetitive calculations. The dynamic programming code using \Eq{eq:recursion2} avoids this shortcoming. In conclusion, this new recursion formula enables analytical calculation of all conditional probabilities (partition functions) with manageable computational cost and sufficient accuracy.

\section{Masked variational autoregressive network}
\label{sec:masked_van}
The variational autoregressive network (VAN)~\cite{PhysRevLett.122.080602} is one of three neural networks employed in the variational free energy method. It is used to model the electron excitation distribution $p(\vec{k}|\vec{s})$ over HF single-particle states $\psi_{\vec{s},k}$ when nuclear coordinates $\vec{s}$ are given.
The VAN we used consists of two main parts: the conditional probabilities of ideal electrons on HF orbitals, and the masked logit generated by the autoregressive model. 
The conditional probability part encodes the distribution of non-interacting electrons on the HF energy levels $\epsilon_{\vec{s},k}$, which serves as a good starting point of $p(\vec{k}|\vec{s})$, particularly as the HF energy levels vary with nuclear motion. This part is analytical and untrainable.
The trainable part is the masked logit. It is used to capture the difference between real many-body electron excitation and ideal electron excitation.

The combination of these two parts is to add the logarithms of two probabilities and then apply the softmax function. We give the conditional probability of $N$-ideal fermion system in \Eq{eq:qk1} and \Eq{eq:qki}. On Hartree-Fock level, considering spin, the $N/2$ spin-up and $N/2$ spin-down electrons subsystems can be treated as two independent $N/2$ ideal fermion systems on the HF orbitals. Note that the conditional probability we used in this section has to change the particle number from $N$ to $N/2$ for each spin. 
When taking the interaction into account, the joint probability of the spin-up and spin-down electrons is no longer independent, although the marginal distributions of two subsystems are still the same due to the spin symmetry of the Hamiltonian. Thus, the probability has the form of \Eq{eq:transformer1}. The change of introducing interaction is represented by the masked logit. The merged conditional probability of the spin-up electrons is given by
\begin{subequations}
    \begin{align}
        p(k_1^{\uparrow}|\vec{s}) &= \textrm{softmax}\left[\textrm{logit}(k_1^{\uparrow})+\ln p_f^{N/2}(k_1^{\uparrow})\right] \, , \\
        p(k_i^{\uparrow}|k_1^{\uparrow},\cdots,k_{i-1}^{\uparrow},\vec{s}) &= \textrm{softmax}\left[\textrm{logit}(k_i^{\uparrow}|k_1^{\uparrow},\cdots,k_{i-1}^{\uparrow})+ \ln p_f^{N/2}(k_i^{\uparrow}|k_{i-1}^{\uparrow})\right] \, .
    \end{align}
\end{subequations}
Similarly, the conditional probability of the spin-down electrons is given by
\begin{subequations}
    \begin{align}
        p(k_1^{\downarrow}|\vec{k}^{\uparrow},\vec{s}) &= \textrm{softmax}\left[\textrm{logit}(k_1^{\downarrow}|\vec{k}^{\uparrow})+\ln p_f^{N/2}(k_1^{\downarrow})\right] \, , \\
        p(k_i^{\downarrow}|k_1^{\downarrow},\cdots,k_{i-1}^{\downarrow},\vec{k}^{\uparrow},\vec{s}) &= \textrm{softmax}\left[\textrm{logit}(k_i^{\downarrow}|k_1^{\downarrow},\cdots,k_{i-1}^{\downarrow},\vec{k}^{\uparrow})+\ln p_f^{N/2}(k_i^{\downarrow}|k_{i-1}^{\downarrow})\right] \, .
    \end{align}
\end{subequations}
The nuclear configuration $\vec{s}$ is incorporated into $p(\vec{k}|\vec{s})$ through the conditional probabilities of ideal fermions, and the interaction between the electrons, including same spin and opposite spin, is captured by the masked logit.
The complete masked VAN output integrates these conditional probabilities into \Eq{eq:transformer1}, \Eq{eq:transformer2a}, and \Eq{eq:transformer2b}, with the softmax values normalized to ensure proper probability distributions.

\begin{figure}[h]
    \centering
    \includegraphics[width=0.8\linewidth]{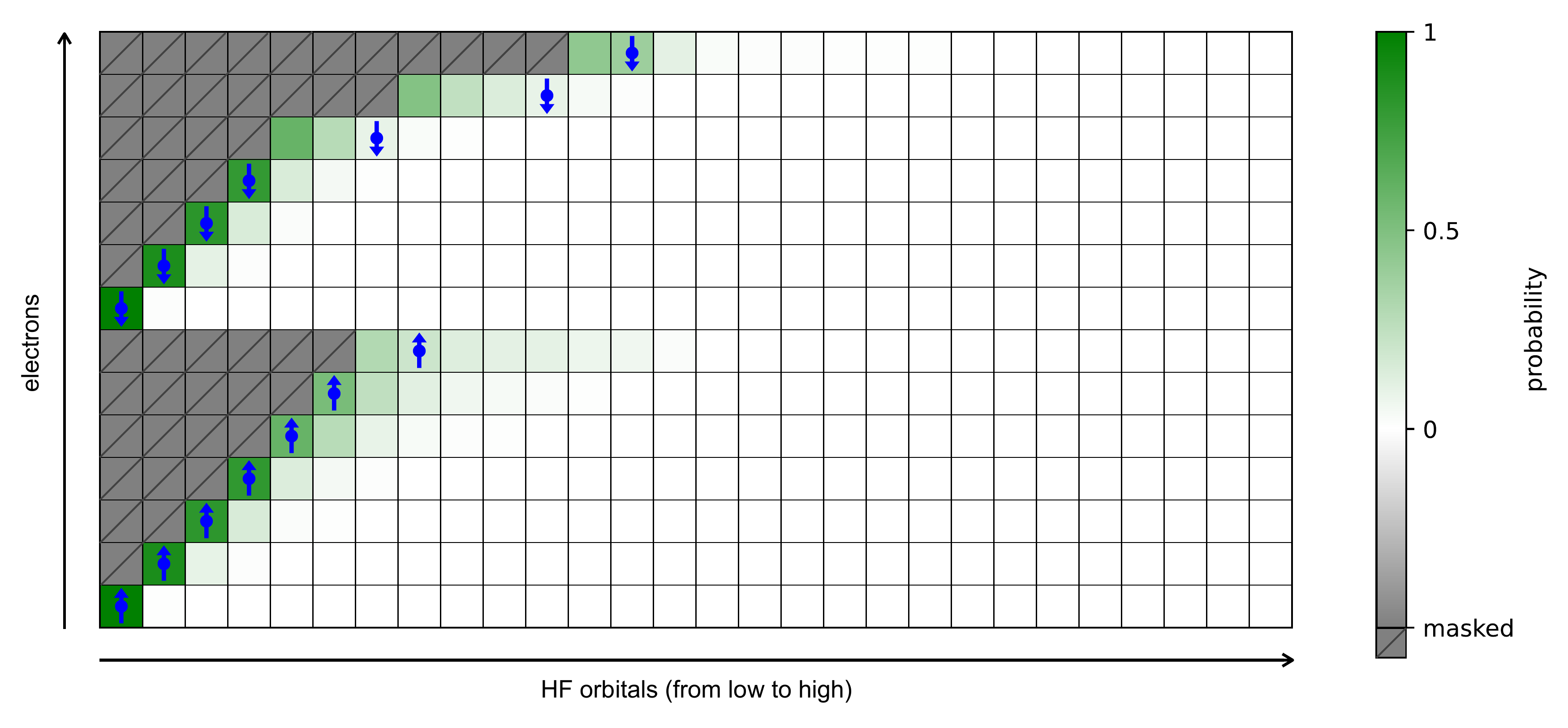}
    \caption{Conditional probability visualization of masked VAN for an electron occupation snapshot. There are 14 rows in the figure, corresponding to $N=14$ electrons. Electron occupancy is calculated from bottom to top in this figure. Each row represents a set of conditional probabilities on $28$ HF orbitals for an electron to occupy. The blue symbols with up/down arrows indicate the spin-up/down electron occupied orbitals, which are sampled from the conditional probabilities. The green and white grids are the areas that can be occupied, while the gray grids with lines indicate the masked area where the electron cannot occupy, depending on the previously occupied state. Darker colors indicate higher probabilities. The probabilities in each row are normalized.
    }
    \label{fig:maskvan}
\end{figure}

The logit value is computed by the masked autoregressive model. The original VAN~\cite{PhysRevLett.122.080602} is implemented using MADE~\cite{pmlr-v37-germain15} and PixelCNN~\cite{pmlr-v48-oord16} to solve the lattice model problem. The VAN we used in the variational free energy method follows the architecture of VAN in neural canonical transformation~\cite{10.21468/SciPostPhys.14.6.154}: an autoregressive transformer~\cite{NIPS2017_3f5ee243}. The algorithm table of masked VAN is in \Alg{alg:maskvan}. The mask ensures that the occupation numbers $\vec{k}$ adhere to the prescribed order $1 \leq k_1^{\sigma} < k_2^{\sigma} < \cdots < k_{N/2}^{\sigma} \leq M$ for spin $\sigma = \uparrow$ or $\downarrow$, in accordance with the Pauli exclusion principle. It also ensures the uniqueness of the probability for one $\vec{k}$. The masking method is the same as in Ref.~\cite{10.21468/SciPostPhys.14.6.154}. The addition of the conditional probability for ideal fermion makes VAN have the same starting point as the pretrained VAN in Ref.~\cite{10.21468/SciPostPhys.14.6.154}, and this is indispensable in case the HF orbitals vary when the nuclei are moving. The sketch of masked conditional probability is shown in \Fig{fig:maskvan}. Electrons occupy the HF orbitals from bottom to top, and the gray masked area ensures that the next occupation can only be on a higher energy orbital than the previous one within the same spin. The distribution of electron occupation on HF orbitals is shown in \Fig{fig:train_van} in \ref{sec:training}.

\begin{algorithm}[H]
	\begin{algorithmic}[1]
		\caption{Masked variational autoregressive network.}
        \Require Hartree-Fock energy levels $\epsilon_{\vec{s},k}$.
        \Ensure Sampled $\vec{k}$ and corresponding log-probability $\ln p(\vec{k}|\vec{s})$.
        \For{$i=1,\cdots ,N/2$} \Comment Spin-up electrons
            \State $\textrm{logit}(k_i^{\uparrow}|k_1^{\uparrow},\cdots,k_{i-1}^{\uparrow}) = {\color{blue}\texttt{Transformer}} (k_1^{\uparrow},\cdots,k_{i-1}^{\uparrow})$
            \State For $k_{i}^{\uparrow} \leq k_{i-1}^{\uparrow}$, set $\textrm{logit}(k_i^{\uparrow}|k_1^{\uparrow},\cdots,k_{i-1}^{\uparrow}) \equiv -\infty $ \Comment Add mask
            \State Sample $k_i^{\uparrow} \sim p(k_i^{\uparrow}|k_1^{\uparrow},\cdots,k_{i-1}^{\uparrow},\vec{s}) = \textrm{softmax}\left[\textrm{logit}(k_i^{\uparrow}|k_1^{\uparrow},\cdots,k_{i-1}^{\uparrow})+ \ln p_f^{N/2}(k_i^{\uparrow}|k_{i-1}^{\uparrow})\right]$
        \EndFor
        \For{$i=1,\cdots ,N/2$} \Comment Spin-down electrons
            \State $\textrm{logit}(k_i^{\downarrow}|k_1^{\downarrow},\cdots,k_{i-1}^{\downarrow},\vec{k}^{\uparrow}) = {\color{blue}\texttt{Transformer}} (k_1^{\downarrow},\cdots,k_{i-1}^{\downarrow},\vec{k}^{\uparrow})$
            \State For $k_{i}^{\downarrow} \leq k_{i-1}^{\downarrow}$, set $\textrm{logit}(k_i^{\downarrow}|k_1^{\downarrow},\cdots,k_{i-1}^{\downarrow},\vec{k}^{\uparrow}) \equiv -\infty$ \Comment Add mask
            \State Sample $k_i^{\downarrow} \sim p(k_i^{\downarrow}|k_1^{\downarrow},\cdots,k_{i-1}^{\downarrow},\vec{k}^{\uparrow},\vec{s}) = \textrm{softmax}\left[\textrm{logit}(k_i^{\downarrow}|k_1^{\downarrow},\cdots,k_{i-1}^{\downarrow},\vec{k}^{\uparrow})+ \ln p_f^{N/2}(k_i^{\downarrow}|k_{i-1}^{\downarrow})\right]$
        \EndFor
        \State $\vec{k} = \{k_1^\uparrow, \cdots, k_{N/2}^\uparrow,$ $k_1^\downarrow, \cdots, k_{N/2}^\downarrow\}$ 
        \State $\ln p(\vec{k}) = \ln p(k_1^{\uparrow}|s)+\cdots+\ln p(k_{N/2}^{\uparrow}|k_1^{\uparrow},\cdots,k_{N/2-1}^{\uparrow},\vec{s}) + \ln p(k_1^{\downarrow}|\vec{k}^{\uparrow},s)+\cdots+\ln p(k_{N/2}^{\downarrow}|k_1^{\downarrow},\cdots,k_{N/2-1}^{\downarrow},\vec{k}^{\uparrow},\vec{s})$
	    \State \Return $\vec{k}$, $\ln p(\vec{k})$
		\label{alg:maskvan}
	\end{algorithmic}
\end{algorithm}

\section{Flow model architectures}
\label{sec:flow}
In the variational free energy method, we employ two flow models to represent the nuclear Boltzmann distribution and the backflow transformation of electrons. Each flow model is a bijective mapping between real particles and quasiparticles, implemented using permutationally equivariant layers from FermiNet~\cite{PhysRevResearch.2.033429}. The architecture of the nucleus flow model, which describes the Boltzmann distribution, is identical to that used in the ground-state variational free energy method~\cite{PhysRevLett.131.126501, rbsg-r7hx}. The algorithm table for the nucleus flow model is shown in \Alg{alg:flow}. The nucleus flow model takes nuclear coordinates $\vec{s}$ and box length $L$ as inputs, and outputs the log-probability of the nuclear configuration $\ln p(\vec{s})$.
\begin{algorithm}[H]
	\begin{algorithmic}[1]
		\caption{Normalizing flow for nuclei Boltzmann distribution.}
        \Require Nuclear coordinates $\vec{s}$ and box length $L$. 
        \Ensure Log-probability $\ln p(\vec{s})$
        \State $\boldsymbol{f} ={\color{blue}  \texttt{FermiNet}} (\boldsymbol{s}, [], L)$   \Comment $\mathbb{R}^{N\times M }  $ 
        \State $\boldsymbol{\zeta} = \boldsymbol{s} + {\color{blue}\texttt{FC}_{3}}(\boldsymbol{f})$ \Comment $\mathbb{R}^{N\times 3}$ 
		\State \Return $ \ln \left|\det
		\left( \frac{\partial \boldsymbol{\zeta} }{\partial \boldsymbol{s}} \right)\right|  -\ln ( L^{3N})$
		\label{alg:flow}
	\end{algorithmic}
\end{algorithm}
The flow model for the electron wave function is the backflow part of the ground-state variational free energy method's wave function. The algorithm table for the electron flow model is shown in \Alg{alg:backflow}. The electron flow model takes nuclear coordinates $\vec{s}$, electron coordinates $\vec{r} = [\vec{r}^\uparrow, \vec{r}^\downarrow]$, and box length $L$ as inputs, and outputs the log-wave function $\ln \Psi_{\vec{s},\vec{k}}(\vec{r})$.
\begin{algorithm}[H]
	\begin{algorithmic}[1]
		\caption{Flow model for the electron wave function.}
        \Require Nuclear coordinates $\boldsymbol{s}$, electron coordinates $\boldsymbol{r} = [ \boldsymbol{r}^\uparrow, \boldsymbol{r}^\downarrow ]$, and box length $L$.
        \Ensure Log-wave function $\ln \Psi_{\vec{s},\vec{k}}(\vec{r})$. 
        \State $\boldsymbol{f}^{\vec{s}}, \boldsymbol{f}^\uparrow, \boldsymbol{f}^\downarrow = {\color{blue}\texttt{FermiNet}} ([\boldsymbol{s} , \boldsymbol{r}^\uparrow,  \boldsymbol{r}^\downarrow], [N, 3N/2], L)$ 
	    \State $\boldsymbol{\xi}^{\uparrow, \downarrow} = \boldsymbol{r}^{\uparrow, \downarrow} + {\color{blue}\texttt{FC}_{3}}(\boldsymbol{f}^{\uparrow, \downarrow} ) $  \Comment Backflow coordinates  
	    \State \Return $\ln \left|\det\left[\psi_{\vec{s},k_i^{\uparrow}}(\boldsymbol{\xi}_j^{\uparrow})\right]\right| + \ln\left|\det\left[\psi_{\vec{s},k_i^{\downarrow}}(\boldsymbol{\xi}_j^{\downarrow})\right]\right| + \frac{1}{2}\ln\left|\det\left(\frac{\partial\boldsymbol{\xi}}{\partial \vec{r}}\right)\right|$ 
		\label{alg:backflow}
	\end{algorithmic}
\end{algorithm}

\section{Training}
\label{sec:training}
Training the neural networks in the variational free energy method involves two primary stages: pretraining the nucleus flow model and joint training of the nucleus flow model, electron flow model, and VAN. Pretraining is a low-cost, rapid process that enables the nucleus flow model to approximate the finite-temperature HF Boltzmann distribution. In contrast, joint training, which is the core of the variational free energy method, is computationally intensive and time-consuming, as it optimizes the three neural networks to learn the many-body density matrix of the system.

\begin{figure}[h]
    \centering
    \includegraphics[width=\linewidth]{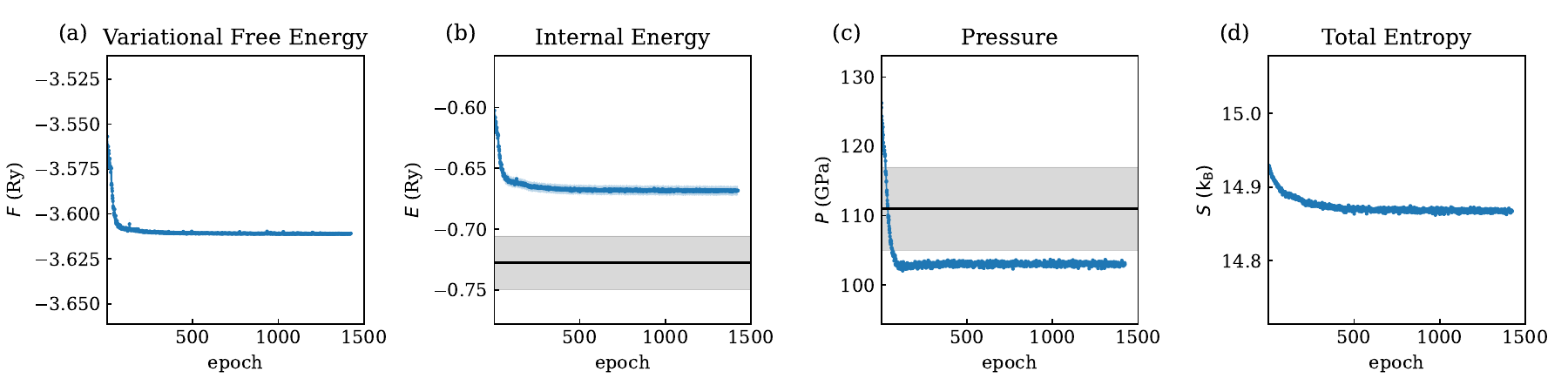}
    \caption{Pretraining curve of the nucleus flow model $p(\vec{s})$ for a TBC deuterium system with $N=32$ at $T=31\,250$ K and $r_s=2$. Panels show (a) variational free energy $F$ (loss function), (b) internal energy $E$, (c) pressure $P$, and (d) total entropy $S$. The physical quantities, except for $P$, are quantities per atom. Energies are in Rydberg units and entropies are in $k_{\mathrm{B}}$ units. The same applies to the following content. The blue line represents the average quantities with error bars (blue-filled area) for a batch size of $400 \times 2$, and the black solid line denotes the benchmark results from RPIMC~\cite{PhysRevLett.85.1890}. The gray-filled areas are the error bars of RPIMC.}
    \label{fig:pretrain}
\end{figure}

The pretraining of the nucleus flow model, $p(\vec{s})$, is an effective step to minimize the variational free energy. We pretrain the model using the HF total energy $E_{\text{HF}}(\vec{s})$ as the potential energy surface to achieve the HF Boltzmann distribution before joint training. The HF PES is also solved by the batched HF solver. The loss function for pretraining is
\begin{equation}
    F = \mathop{\mathbb{E}}_{\vec{s}\sim p(\vec{s})} \left[\frac{1}{\beta}\ln{p(\vec{s})}
    -\frac{1}{\beta}S_{\text{HF}}(\vec{s})+E_{\text{HF}}(\vec{s})\right] \, ,
\end{equation}
where $S_{\text{HF}}(\vec{s})$ is the finite-temperature HF electronic entropy. We employ stochastic reconfiguration~\cite{Becca_Sorella_2017} to pretrain the flow model, using Fermi-Dirac smearing in the HF calculations, consistent with the initial VAN probability distribution. Each pretraining epoch comprises four steps: (1) sampling $\vec{s}$ from the nucleus distribution $p(\vec{s})$, (2) computing the HF PES, (3) calculating the loss and gradient, and (4) updating the network parameters. The training curve for the pretraining of the nucleus flow model is shown in \Fig{fig:pretrain}. After pretraining, the variational free energy and internal energy converge to the finite-temperature HF molecular dynamics results. In the subsequent joint training, we set the learning rate of the nucleus flow model to $0.1 \sim 0.3$, smaller than that of the electron flow model and VAN (set to 1.0), as the pretrained nucleus flow model is already near the optimal solution and requires only fine-tuning.

\begin{figure}[h]
    \centering
    \includegraphics[width=\linewidth]{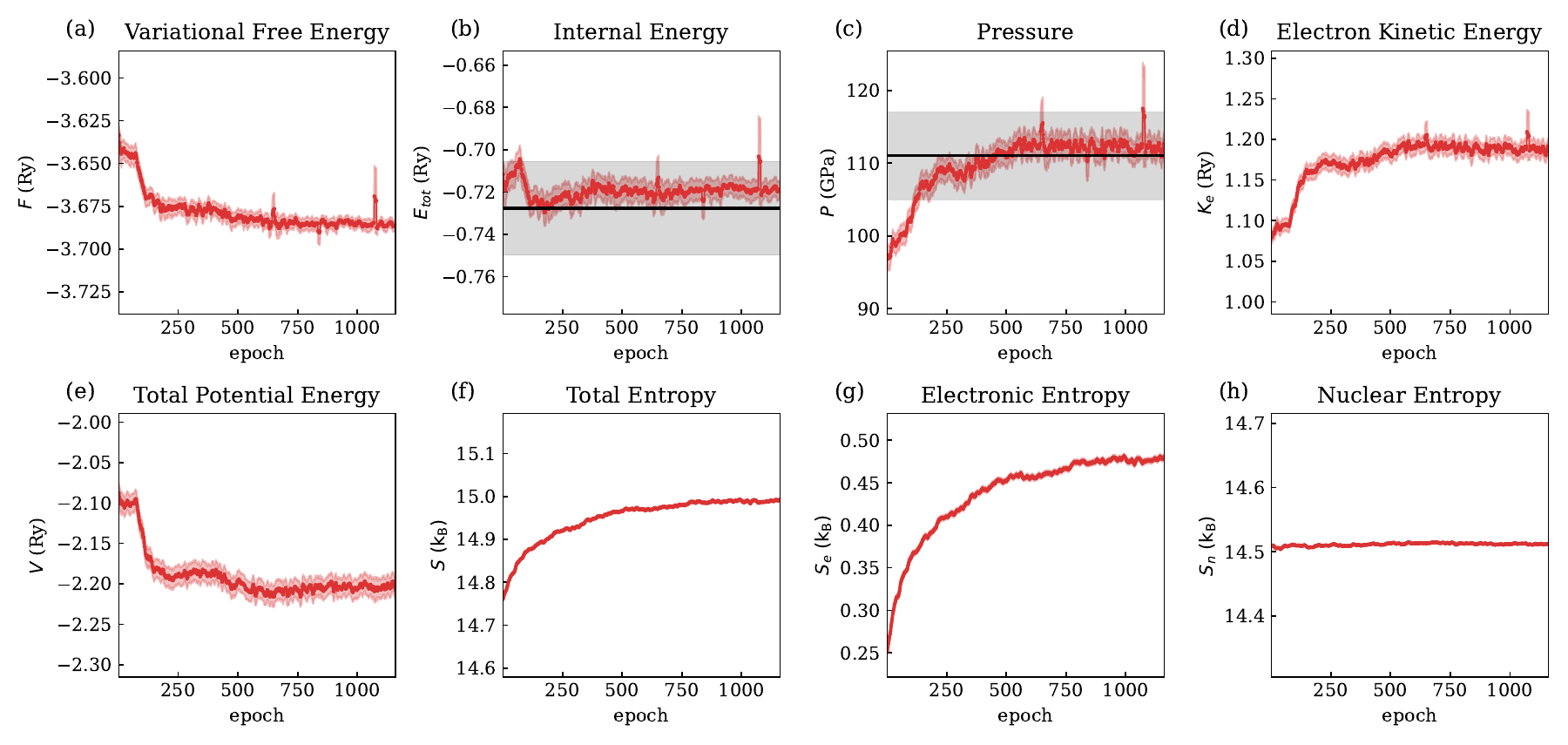}
    \caption{Training curve of the variational free energy method for a TBC deuterium system with $N=32$ at $T=31\,250$ K and $r_s=2$. Panels show (a) variational free energy $F$ (loss function), (b) internal energy $E$, (c) pressure $P$, (d) electron kinetic energy $K_e$, (e) total potential energy $V = V_{ee} + V_{en} + V_{nn}$, (f) total entropy $S$, (g) electronic entropy $S_e$, and (h) nuclear entropy $S_n$. Panels (a), (b), and (c) are the same as the one in the main text. The red line represents the average quantities with error bars (red-filled area) for a batch size of $320 \times 8$. The starting point in this figure used the pretrained flow model from the final checkpoint in \Fig{fig:pretrain}. The black solid line denotes the PBC benchmark results from RPIMC~\cite{PhysRevLett.85.1890}. The gray-filled areas are the error bars of RPIMC.}
    \label{fig:train}
\end{figure}

Joint training is the cornerstone of the variational free energy method, optimizing the nucleus flow model, electron flow model, and VAN by minimizing the variational free energy in \Eq{eq:FreeEnergy}. Each training epoch involves six steps: (1) sampling $\vec{s}$ from $p(\vec{s})$, (2) computing HF orbitals and energy levels, (3) sampling $\vec{k}$ from $p(\vec{k}|\vec{s})$, (4) sampling from the electron wave function $|\Psi_{\vec{s},\vec{k}}(\vec{r})|^2$, (5) calculating the loss and gradients, and (6) updating the network parameters. For a deuterium system with $N=20$ or $N=32$, the batch size is typically $320 \times 8$, where 320 is the base batch size, and the factor of 8 represents an effective increase achieved through independent samples obtained via Markov chain Monte Carlo (MCMC) steps. The training curve for joint training is shown in \Fig{fig:train}.

\begin{figure}[h]
    \centering
    \includegraphics[width=0.3\linewidth]{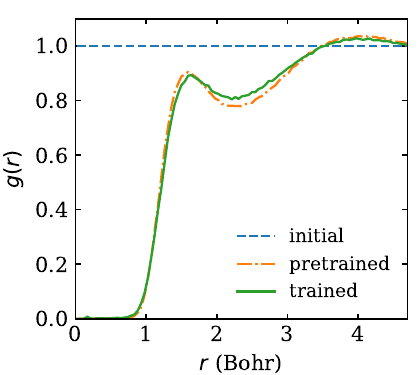}
    \caption{The nucleus-nucleus RDF for PBC system at $N=32$, $T=10\,000$ K, and $r_s=1.86$. The blue line is the theoretical RDF of the initial flow model $p(\vec{s})$. The orange line is the RDF sampled from the pretrained flow model. The green line is the RDF sampled from the joint trained flow model.}
    \label{fig:train_rdf}
\end{figure}

To visualize the three neural networks in the training process, we can draw samples from the three probability distributions. We use radial distribution functions (RDF) to visualize the samples from the flow model probability distribution $p(\vec{s})$. At the beginning of pretraining, shown in \Fig{fig:train_rdf}, the nucleus flow model is almost an identity transformation, and the initial nucleus distribution is uniform. After pretraining, the nucleus flow model learns the HF PES. The probability of nuclei being distributed in close proximity clearly decreases, and the molecular peak emerges under some conditions (short $r_s$ and low temperature). After joint training, the nucleus distribution is further optimized, and the RDF shifts a little from the pretrained RDF. 

\begin{figure}[h]
    \centering
    \includegraphics[width=0.4\linewidth]{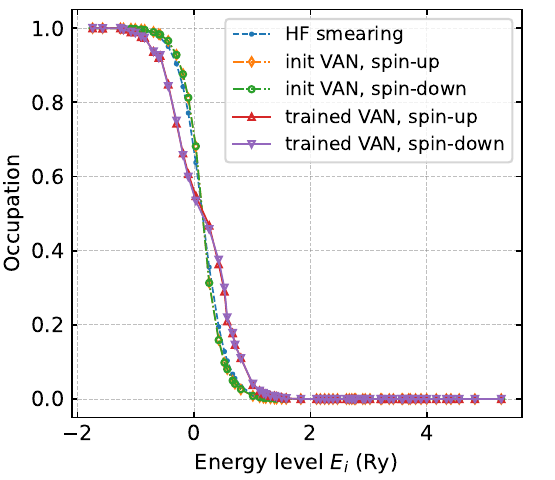}
    \caption{The electron distribution on HF energy levels for a single $N=32$, $T=31\,250$ K, and $r_s=1.86$ PBC nuclear configuration. The blue line is the HF smearing distribution of the HF energy levels. The orange and green lines are the spin-up and spin-down electron distribution sampled from the initial VAN $p(\vec{k}|\vec{s})$. The red and purple lines are the spin-up and spin-down electron distribution sampled from the trained VAN.}
    \label{fig:train_van}
\end{figure}

As for VAN, since the output layer of the initial VAN is a linear layer with weights initialized to have small variance, the initial logits are nearly zero. As a result, the initial VAN distribution is close to Fermi-Dirac smearing of the HF energy levels. After training, the distribution of VAN becomes flatter, indicating that the electron excitation is more uniform across the orbitals. The sampled electron occupation on HF orbitals for a single nuclear configuration is shown in \Fig{fig:train_van}. The figure also shows that the spin-up and spin-down electron distributions are almost the same, which is consistent with the spin symmetry of the Hamiltonian.

The electron flow model is similar to the nucleus flow model. It is an identity transformation at the beginning of joint training, and it learns the backflow transformation during the training process. It is hard to visualize this directly, since this transformation is in the HF wave function. One can turn to the nucleus-electron and electron-electron RDFs in \ref{sec:rdf} to see the transformed electron distribution.

\section{Data analysis}
\label{sec:dataprocessing}
After training the networks, we sample extensively ($\geq 50\,000$ samples) from the optimized distributions to compute the EOS. This process, which constitutes the inference phase of the trained networks, mirrors the training process but omits gradient computation and parameter update. The inference process involves five steps: (1) sampling $\vec{s}$ from the nucleus distribution $p(\vec{s})$, (2) computing Hartree-Fock (HF) orbitals and energy levels, (3) sampling $\vec{k}$ from $p(\vec{k}|\vec{s})$, (4) sampling from the electron wave function $\left|\Psi_{\vec{s},\vec{k}}(\vec{r})\right|^2$, and (5) calculating physical quantities. The explicit expressions for the nucleus distribution and electron excitation probability distribution enable direct evaluation of nuclear and electronic entropies. The internal energy comprises three components: (1) the electron kinetic energy, which is computed via automatic differentiation, (2) the potential energy, which is calculated using Ewald summation, and (3) the nuclear kinetic energy, which is given by $\frac{3}{2} k_{\mathrm{B}} T$. The pressure is determined using the virial theorem
\begin{equation}
   3P\Omega = 2K+V \, ,\label{eq:virial}
\end{equation}
where $P$ is the pressure, $\Omega$ is the volume per atom, $K$ and $V$ are the kinetic and potential energies per atom, respectively. Note that the nuclear entropy has to consider the contribution of the integral over the momenta, and the only difference between hydrogen and deuterium in the variational free energy method is that the mass of the nucleus is different here; for details, see the supplemental material of Ref.~\cite{PhysRevLett.131.126501}. The EOS results of TBC and PBC systems without finite-size correction are presented in \Table{tab:EOS} and \Table{tab:EOS_PBC}, respectively. 
Part of the data in \Table{tab:EOS} is the raw data without finite-size correction corresponding to \Table{tab:eos} in the main text.
Note that the quantities, except for $F$ and $S_n$ in the table, also work for hydrogen.

\begin{table}[H]
    \centering
    \renewcommand{\arraystretch}{1.2}
    \resizebox{\textwidth}{!}{
    \begin{tabular}{ c c c c c c c c c c c c }
        \hline \hline
        $N$ & $T\ (\mathrm{K})$ & $r_s$ & $F\ (\mathrm{Ry})$ & $E\ (\mathrm{Ry})$ & $V_{en}\ (\mathrm{Ry})$ & $V_{ee}\ (\mathrm{Ry})$ & $V_{nn}\ (\mathrm{Ry})$ & $K_e\ (\mathrm{Ry})$ & $P\ (\mathrm{GPa})$ & $S_e\ (k_{\mathrm{B}})$ & $S_n\ (k_{\mathrm{B}})$ \\
        \hline
        14 & 10000 & 1.86 & -1.7305(16) & -0.9960(16) & -1.0467(26) & -0.53275(75) & -0.6859(14) & 1.1744(24) & 49.74(60) & 0.04325(31) & 11.55387(69) \\
        20 & 10000 & 1.86 & -1.7367(18) & -0.9773(17) & -1.1489(27) & -0.49821(80) & -0.6455(16) & 1.2203(24) & 61.50(65) & 0.05936(30) & 11.93074(62) \\
        32 & 10000 & 1.86 & -1.77445(61) & -0.98405(61) & -1.1689(17) & -0.47903(55) & -0.64540(23) & 1.2143(16) & 59.17(34) & 0.05493(22) & 12.42446(38) \\
        32 & 10000 & 2.00 & -1.79347(63) & -0.98779(63) & -1.2895(18) & -0.41229(54) & -0.54991(26) & 1.1689(17) & 40.41(28) & 0.07003(25) & 12.65063(42) \\
        32 & 15625 & 1.86 & -2.23601(51) & -0.91661(51) & -1.2044(18) & -0.46388(55) & -0.61879(27) & 1.2220(17) & 82.57(33) & 0.16002(35) & 13.17234(28) \\
        32 & 15625 & 2.00 & -2.2627(17) & -0.9178(17) & -1.3024(18) & -0.39856(55) & -0.53035(30) & 1.1651(21) & 57.90(51) & 0.17514(37) & 13.41464(28) \\
        32 & 31250 & 1.86 & -3.63514(60) & -0.72231(61) & -1.2669(18) & -0.43510(60) & -0.57470(35) & 1.2575(16) & 151.38(33) & 0.43886(51) & 14.27798(21) \\
        32 & 31250 & 2.00 & -3.68644(49) & -0.71923(51) & -1.3290(18) & -0.37531(58) & -0.50516(35) & 1.1934(16) & 112.82(26) & 0.47941(51) & 14.51214(19) \\
        32 & 62500 & 1.86 & -6.7116(37) & -0.3061(37) & -1.3063(21) & -0.41141(63) & -0.53356(42) & 1.3514(36) & 298.2(13) & 0.83528(60) & 15.34620(18) \\
        32 & 62500 & 2.00 & -6.81638(61) & -0.29995(66) & -1.3380(19) & -0.36366(60) & -0.47793(41) & 1.2859(17) & 231.15(28) & 0.88904(61) & 15.57281(17) \\
        54 & 10000 & 1.86 & -1.7930(18) & -0.9741(18) & -1.2406(34) & -0.4471(11) & -0.61036(54) & 1.2290(31) & 63.66(71) & 0.02195(42) & 12.90718(86) \\
        54 & 10000 & 2.00 & -1.8069(13) & -0.9737(13) & -1.3396(34) & -0.3825(11) & -0.52503(55) & 1.1784(30) & 43.85(49) & 0.02707(36) & 13.12854(94) \\
        \hline \hline
    \end{tabular}
    }
    \caption{The variational free energy equation of state for TBC deuterium system without finite-size correction. The quantities in the table are particle number $N$, temperature $T$, Wigner-Seitz parameter $r_s$, internal energy $E$, electron-nucleus potential energy $V_{en}$, electron-electron potential energy $V_{ee}$, nucleus-nucleus potential energy $V_{nn}$, electron kinetic energy $K_e$, pressure $P$, electronic entropy $S_e$, nuclear entropy $S_n$. All of the energies and entropies in this table are energies per atom of the $N$-atom system.}
    \label{tab:EOS}
\end{table}

\begin{table}[H]
    \centering
    \renewcommand{\arraystretch}{1.2}
    \resizebox{\textwidth}{!}{
    \begin{tabular}{ c c c c c c c c c c c c }
        \hline \hline
        $N$ & $T\ (\mathrm{K})$ & $r_s$ & $F\ (\mathrm{Ry})$ & $E\ (\mathrm{Ry})$ & $V_{en}\ (\mathrm{Ry})$ & $V_{ee}\ (\mathrm{Ry})$ & $V_{nn}\ (\mathrm{Ry})$ & $K_e\ (\mathrm{Ry})$ & $P\ (\mathrm{GPa})$ & $S_e\ (k_{\mathrm{B}})$ & $S_n\ (k_{\mathrm{B}})$ \\
        \hline
        14 & 10000 & 1.86 & -1.72235(80) & -0.98901(80) & -1.1922(25) & -0.48713(74) & -0.63798(32) & 1.2333(23) & 61.72(45) & 0.03565(29) & 11.54293(67) \\
        14 & 10000 & 2.00 & -1.73408(91) & -0.98802(91) & -1.3455(24) & -0.41250(75) & -0.53502(35) & 1.2100(22) & 46.39(37) & 0.000025(14) & 11.77931(67) \\
        14 & 15625 & 1.86 & -2.1229(13) & -0.9022(13) & -1.2733(25) & -0.46703(74) & -0.60917(38) & 1.2988(26) & 99.15(62) & 0.15078(41) & 12.18407(54) \\
        14 & 15625 & 2.00 & -2.16814(80) & -0.91628(80) & -1.3457(25) & -0.40753(76) & -0.52948(40) & 1.2180(23) & 65.87(38) & 0.11542(46) & 12.53436(50) \\
        14 & 31250 & 1.86 & -3.42207(94) & -0.73971(93) & -1.2696(25) & -0.46257(80) & -0.59021(52) & 1.2858(23) & 153.35(48) & 0.10037(61) & 13.45201(53) \\
        14 & 31250 & 2.00 & -3.4926(10) & -0.7503(10) & -1.3121(25) & -0.41550(74) & -0.53548(42) & 1.2159(25) & 111.59(44) & 0.16102(66) & 13.69439(25) \\
        14 & 62500 & 1.86 & -6.36571(78) & -0.30682(85) & -1.2646(26) & -0.45536(81) & -0.58050(48) & 1.3999(25) & 306.87(52) & 0.77980(86) & 14.52622(23) \\
        14 & 62500 & 2.00 & -6.4779(21) & -0.3082(21) & -1.2918(26) & -0.40765(75) & -0.52626(46) & 1.3237(28) & 235.48(64) & 0.83484(85) & 14.75093(21) \\
        20 & 10000 & 1.50 & -1.6428(26) & -0.9295(26) & -1.0273(16) & -0.64887(54) & -0.89329(18) & 1.5449(27) & 246.4(17) & 0.04806(13) & 11.21360(37) \\
        20 & 10000 & 1.75 & -1.7189(27) & -0.9717(27) & -1.1957(17) & -0.51658(50) & -0.69006(21) & 1.3357(34) & 100.3(13) & 0.04907(19) & 11.74792(36) \\
        20 & 10000 & 1.86 & -1.74380(66) & -0.98295(66) & -1.2584(16) & -0.46145(49) & -0.61271(22) & 1.2546(15) & 66.70(31) & 0.06642(22) & 11.94646(38) \\
        20 & 10000 & 1.94 & -1.75219(60) & -0.98724(60) & -1.3176(16) & -0.42445(49) & -0.55935(22) & 1.2192(14) & 52.42(25) & 0.01711(16) & 12.06038(42) \\
        20 & 10000 & 2.00 & -1.76266(43) & -0.98846(43) & -1.3489(16) & -0.40328(48) & -0.52597(23) & 1.1947(14) & 44.08(22) & 0.04166(17) & 12.18196(39) \\
        20 & 15625 & 1.50 & -2.0582(13) & -0.8590(13) & -1.0699(17) & -0.63807(55) & -0.86590(21) & 1.5665(22) & 296.9(11) & 0.13565(28) & 11.98243(29) \\
        20 & 15625 & 1.75 & -2.16010(52) & -0.90839(52) & -1.1999(16) & -0.50463(52) & -0.67664(22) & 1.3243(15) & 123.27(38) & 0.13929(30) & 12.50902(26) \\
        20 & 15625 & 1.86 & -2.18806(48) & -0.91405(48) & -1.2590(16) & -0.45670(50) & -0.60203(24) & 1.2552(15) & 89.08(31) & 0.15933(32) & 12.71440(25) \\
        20 & 15625 & 1.94 & -2.2042(11) & -0.9180(11) & -1.2986(16) & -0.42193(50) & -0.55683(25) & 1.2109(20) & 70.77(46) & 0.14740(31) & 12.84982(26) \\
        20 & 15625 & 2.00 & -2.21744(64) & -0.92214(64) & -1.3310(13) & -0.39962(42) & -0.52368(21) & 1.1837(13) & 60.00(25) & 0.14830(27) & 12.94046(22) \\
        20 & 31250 & 1.50 & -3.33290(75) & -0.67485(76) & -1.1316(17) & -0.61976(58) & -0.81939(28) & 1.5991(17) & 423.54(75) & 0.32603(43) & 13.10353(24) \\
        20 & 31250 & 1.75 & -3.48751(47) & -0.71173(48) & -1.2304(16) & -0.49026(53) & -0.64155(28) & 1.3536(15) & 205.04(37) & 0.40659(48) & 13.61777(19) \\
        20 & 31250 & 1.86 & -3.53995(41) & -0.71833(42) & -1.2586(16) & -0.44594(51) & -0.58539(28) & 1.2747(15) & 155.22(29) & 0.44335(47) & 13.81263(18) \\
        20 & 31250 & 1.94 & -3.57345(55) & -0.72395(56) & -1.2943(17) & -0.41462(50) & -0.54447(29) & 1.2326(15) & 129.15(27) & 0.44654(49) & 13.95035(17) \\
        20 & 31250 & 2.00 & -3.59361(68) & -0.71937(69) & -1.3118(16) & -0.39383(50) & -0.51700(29) & 1.2063(16) & 114.70(29) & 0.47526(50) & 14.04658(16) \\
        20 & 62500 & 1.50 & -6.15209(53) & -0.26434(57) & -1.1759(18) & -0.60067(60) & -0.76315(34) & 1.6817(16) & 697.55(63) & 0.67634(54) & 14.19734(20) \\
        20 & 62500 & 1.75 & -6.42668(56) & -0.30447(60) & -1.2378(17) & -0.48293(54) & -0.61130(33) & 1.4337(16) & 376.36(40) & 0.77890(54) & 14.68707(17) \\
        20 & 62500 & 1.86 & -6.52216(67) & -0.30739(70) & -1.2554(17) & -0.43763(53) & -0.56112(33) & 1.3530(15) & 298.24(33) & 0.82153(56) & 14.87826(16) \\
        20 & 62500 & 1.94 & -6.58546(50) & -0.30801(55) & -1.2712(16) & -0.41023(51) & -0.52777(33) & 1.3074(15) & 255.44(28) & 0.84812(55) & 15.01001(15) \\
        20 & 62500 & 2.00 & -6.63074(45) & -0.30840(50) & -1.2871(16) & -0.38813(52) & -0.50379(33) & 1.2768(15) & 228.60(24) & 0.86528(55) & 15.10627(14) \\
        32 & 10000 & 1.86 & -1.77966(61) & -0.98913(61) & -1.1735(18) & -0.47085(54) & -0.63946(24) & 1.1997(17) & 55.59(35) & 0.06456(24) & 12.41704(39) \\
        32 & 10000 & 1.94 & -1.78698(57) & -0.98994(57) & -1.2629(20) & -0.42723(61) & -0.57315(67) & 1.1783(17) & 45.43(30) & 0.03981(20) & 12.54439(50) \\
        32 & 10000 & 2.00 & -1.79618(57) & -0.98994(57) & -1.2959(18) & -0.40582(53) & -0.54184(26) & 1.1586(16) & 38.58(27) & 0.08174(26) & 12.64777(43) \\
        32 & 15625 & 1.86 & -2.24090(67) & -0.92444(67) & -1.2126(18) & -0.45806(56) & -0.61450(28) & 1.2122(17) & 79.36(37) & 0.13559(36) & 13.16705(29) \\
        32 & 15625 & 2.00 & -2.26600(55) & -0.92071(55) & -1.2982(18) & -0.39504(56) & -0.53110(30) & 1.1551(16) & 56.02(27) & 0.18074(39) & 13.41311(28) \\
        32 & 31250 & 1.86 & -3.63702(52) & -0.72603(53) & -1.2528(18) & -0.43629(60) & -0.57615(35) & 1.2423(16) & 147.93(32) & 0.43249(50) & 14.27500(21) \\
        32 & 31250 & 2.00 & -3.68982(58) & -0.72729(59) & -1.3245(18) & -0.37601(57) & -0.50503(35) & 1.1814(17) & 109.89(27) & 0.45866(55) & 14.50926(19) \\
        32 & 62500 & 1.86 & -6.71197(59) & -0.31176(63) & -1.3005(19) & -0.41027(63) & -0.53785(41) & 1.3430(17) & 295.63(34) & 0.82346(60) & 15.34480(18) \\
        32 & 62500 & 2.00 & -6.8218(19) & -0.3082(19) & -1.3352(19) & -0.35448(62) & -0.47698(42) & 1.2647(22) & 226.85(55) & 0.88223(61) & 15.57255(17) \\
        \hline \hline
    \end{tabular}
    }
    \caption{The variational free energy equation of state for PBC deuterium system without finite-size correction. The quantities in the table are particle number $N$, temperature $T$, Wigner-Seitz parameter $r_s$, internal energy $E$, electron-nucleus potential energy $V_{en}$, electron-electron potential energy $V_{ee}$, nucleus-nucleus potential energy $V_{nn}$, electron kinetic energy $K_e$, pressure $P$, electronic entropy $S_e$, nuclear entropy $S_n$. All of the energies and entropies in this table are energies per atom of the $N$-atom system.}
    \label{tab:EOS_PBC}
\end{table}

\begin{figure}[h]
    \centering 
    \includegraphics[width=0.45\linewidth]{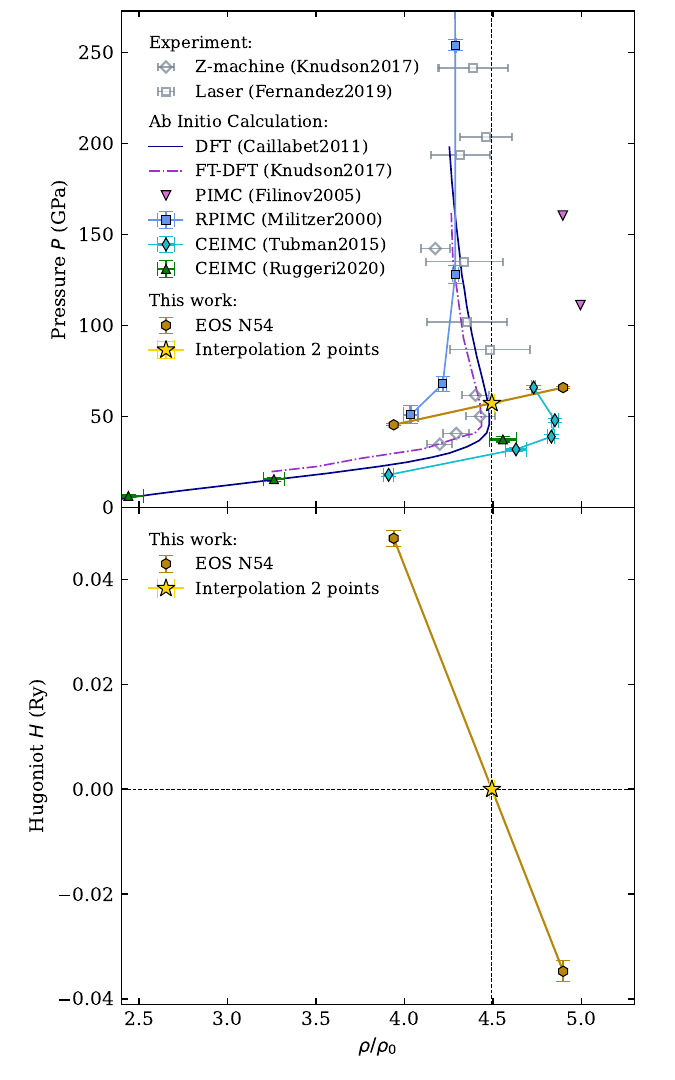}
    \caption{Lagrange interpolation for the Hugoniot curve of a TBC deuterium system with $N=54$ at $10\,000$ K. The upper panel shows the interpolation of the pressure $P_{\textrm{Hugoniot}}$ at density $\rho_{\textrm{Hugoniot}}$, and the lower panel depicts the interpolation of the Hugoniot function $H$. The two dark yellow data points correspond to the EOS at $r_s = 1.86, 2$ from \Table{tab:eos} with finite-size correction. The yellow star indicates the Lagrange interpolation result.}
    \label{fig:interpolation}
\end{figure}

We employ the Lagrange interpolation method to determine points on the deuterium Hugoniot curve, as shown in \Fig{fig:hugoniot}. These points satisfy the Rankine-Hugoniot relation given by \Eq{eq:Hugoniot}. To obtain a single point on the Hugoniot curve from \Table{tab:eos}, we follow two steps. First, we extract the pressure $P$ and internal energy $E$ at a given temperature $T$ from \Table{tab:eos} and calculate the Hugoniot function $H$ at that point. Since the Hugoniot curve typically lies between $r_s = 1.86$ and $r_s = 2$, we select two points at $r_s = 1.86,\ 2$ (or three points at $r_s = 1.86,\ 1.94,\ 2$ under PBC) for a fixed temperature. Using Lagrange interpolation, we determine the density $\rho_{\mathrm{Hugoniot}}$ (or equivalently, the volume $\Omega_{\mathrm{Hugoniot}}$) where $H = 0$, as illustrated in the lower panel of \Fig{fig:interpolation}. In the second step, we apply Lagrange interpolation to compute the pressure $P_{\mathrm{Hugoniot}}$ at the density $\rho_{\mathrm{Hugoniot}}$ on the Hugoniot curve, as shown in the upper panel of \Fig{fig:interpolation}. 
We show a two-point interpolation $N=54$ TBC case. And we get other points on $N=32$ Hugoniot curve at $T=10\,000,\ 15\,625,\ 31\,250,\ 62\,500$ K in the same way. It can be easily verified using the data in \Table{tab:EOS_PBC} that the linear dependence of EOS within this phase diagram region makes the Hugoniot curves obtained by two-point interpolation and three-point interpolation almost the same.

\section{P-T Hugoniot curve}
\begin{figure}[H]
    \centering
    \includegraphics[width=0.45\linewidth]{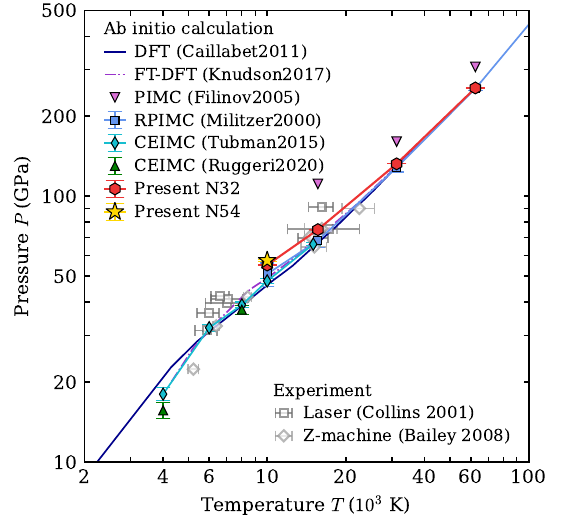}
    \caption{Temperature-pressure diagram of Hugoniot experiments and computations. The red symbols are Lagrange interpolation of variational free energy results of $N=32$ system, , the yellow star is the result at $10\,000$ K for the $N=54$ system. The experimental and theoretical results are from Nova laser~\cite{PhysRevLett.87.165504}, Z-machine~\cite{PhysRevB.78.144107}, DFT~\cite{PhysRevB.83.094101}, FT-DFT~\cite{PhysRevLett.118.035501}, direct PIMC~\cite{filinov_calculation_2005}, RPIMC~\cite{PhysRevLett.85.1890}, and CEIMC~\cite{PhysRevLett.115.045301,PhysRevB.102.144108}.
    }
    \label{fig:hugoniot_tp}
\end{figure}

The pressure-temperature Hugoniot curve is presented in \Fig{fig:hugoniot_tp}. All the points in \Fig{fig:hugoniot_tp} are obtained from \Table{tab:eos} in the main text. The red symbols represent the Lagrange interpolation results of the $N=32$ system, while the yellow star indicates the result at $10\,000$ K for the $N=54$ system. 
The Hugoniot pressure obtained at low temperatures is slightly higher than that predicted by other theoretical approaches, consistent with the experimental measurements~\cite{PhysRevLett.87.165504,PhysRevB.78.144107}. At higher temperatures, our results again agree well with those from RPIMC.

\section{Finite-size effect}
We perform simulations with $N = 14,\ 20,\ 32,\ 54$ atoms under PBC and TBC, with and without finite-size correction at $T=10\,000$ K and $r_s=1.86$, where finite-size effects are expected to be most pronounced, to benchmark finite-size effects.
The finite-size correction we used is the same as Ref.~\cite{PhysRevE.81.021202}.
The leading-order correction to the potential energy per particle is given by
\begin{equation}
    \delta V = \frac{3}{2Nr_s^3}\lim_{\vec{k}\rightarrow 0}\frac{S(\vec{k})}{k^2} \, ,
\end{equation}
where $S(\vec{k})$ is the static structure factor. We use the average value of $S(\vec{k})$ at the smallest non-zero wave vectors to estimate $\lim_{\vec{k}\rightarrow 0}S(\vec{k})/k^2$. The correction to the kinetic energy per particle is based on the random phase approximation (RPA)~\cite{PhysRevE.81.021202}:
\begin{equation}
     \delta V = \frac{\sqrt{3}}{4Nr_s^{3/2}} -  \frac{5.264}{2\pi r_s^2(2N)^{4/3}} \, .
\end{equation}

\begin{figure}[H]
    \centering
    \includegraphics[width=0.65\linewidth]{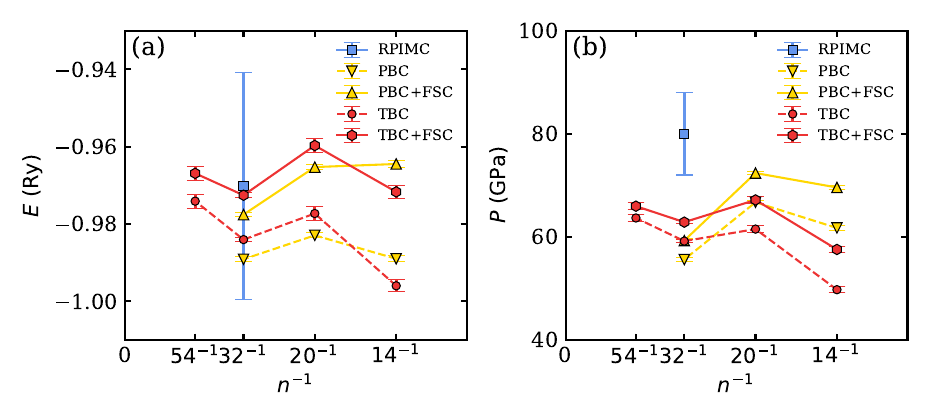}
    \caption{Finite-size effect of variational free energy calculation for deuterium system at $T=10\,000$ K and $r_s=1.86$. 
    The yellow points are results of PBC systems, and the red points are results of TBC systems.
    The dashed lines are raw EOS results without finite-size correction, and the solid lines are EOS results with finite-size correction.
    The blue symbols are RPIMC results~\cite{PhysRevLett.85.1890} of $N=32$ PBC system with VDM nodes.
    }
    \label{fig:fse}
\end{figure}

The finite-size effect of EOS is shown in \Fig{fig:fse}. When PBC is used and no finite-size correction is applied, the finite-size effect is apparent. TBC reduces the finite-size effect to some extent. After applying the finite-size correction, the TBC results show better consistency as the system size increases. The result of $N=54$ system demonstrates the extent of the finite-size effect of EOS in our calculations. The finite-size effect of the Hugoniot curve is shown in \Fig{fig:hugoniot} and \Fig{fig:hugoniot_tp}. The $T=10\,000$ K point on Hugoniot curve of $N=32$ and $N=54$ systems almost overlap, indicating that the finite-size effect on the Hugoniot curve is converged with respect to system size $N$.

\section{Radial distribution functions}
\label{sec:rdf}
\begin{figure}[H]
    \centering
    \includegraphics[width=0.66\linewidth]{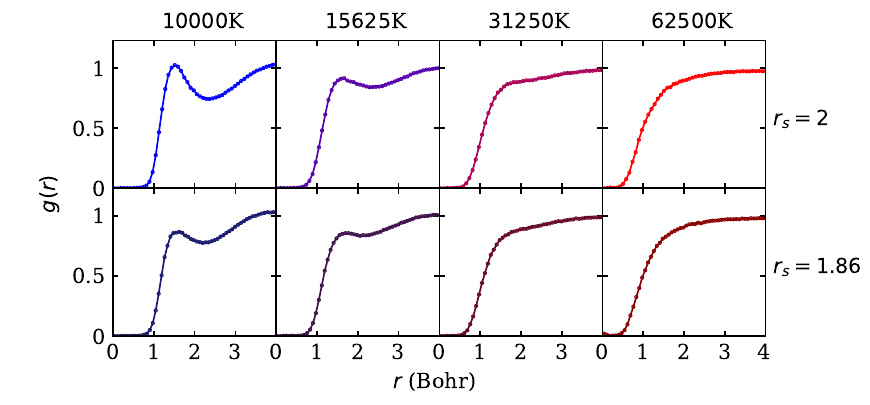}
    \caption{The nucleus-nucleus RDFs at several different temperatures $T$ and $r_s$ of $N=32$ system.}
    \label{fig:rdf_pp}
\end{figure}

\begin{figure}[H]
    \centering
    \includegraphics[width=0.66\linewidth]{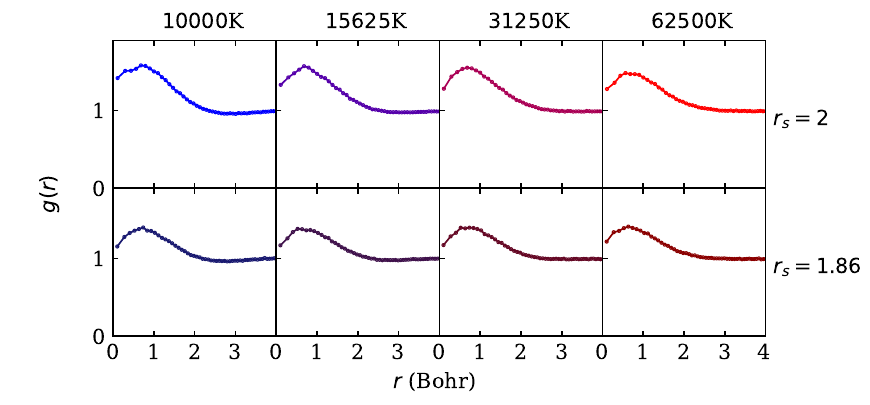}
    \caption{The electron-electron with opposite spin RDFs at several different temperatures $T$ and $r_s$ of $N=32$ system.}
    \label{fig:rdf_eeo}
\end{figure}

\begin{figure}[H]
    \centering
    \includegraphics[width=0.66\linewidth]{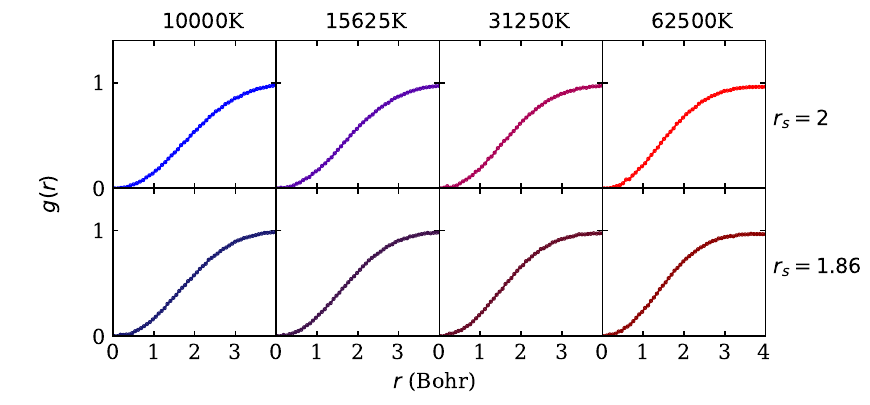}
    \caption{The electron-electron with parallel spin RDFs at several different temperatures $T$ and $r_s$ of $N=32$ system.}
    \label{fig:rdf_eep}
\end{figure}

\begin{figure}[H]
    \centering
    \includegraphics[width=0.66\linewidth]{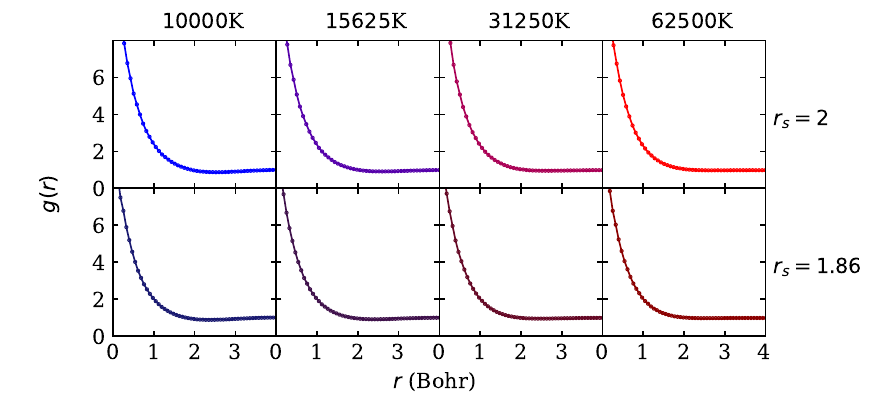}
    \caption{The nucleus-electron RDFs at several different temperatures $T$ and $r_s$ of $N=32$ system.}
    \label{fig:rdf_pe}
\end{figure}

To further benchmark our results and demonstrate the phase transition between atomic and molecular phases, we plot the nucleus-nucleus RDFs of $N=32$ TBC systems at various $r_s$ and temperatures in \Fig{fig:rdf_pp}. As shown in the figure, a clear phase transition from the molecular to the atomic phase occurs when the temperature is increased or the volume is reduced, indicated by the disappearance of the molecular peak. The opposite-spin electron-electron RDFs in \Fig{fig:rdf_eeo} exhibit a peak at short distances, originating from the formation of molecular bonds~\cite{PhysRevE.63.066404}. This peak diminishes as the system becomes sparser or the temperature increases. In contrast, the parallel-spin electron RDFs in \Fig{fig:rdf_eep} show that electrons come progressively closer as $r_s$ decreases, and the RDFs decay to zero at short distances due to the Pauli exclusion principle. The nucleus-electron RDFs in \Fig{fig:rdf_pe} decay exponentially near the nucleus, and become steeper as the system becomes denser. Overall, the RDF results are consistent with those obtained from the variational density matrix (VDM) method~\cite{PhysRevE.61.3470}.

\bibliographystyle{apsrev4-2}
\bibliography{refs}